\documentclass[onecolumn,aps,longbibliography]{revtex4-2}
\usepackage[T1]{fontenc}
\usepackage[latin9]{inputenc}
\usepackage{xcolor,times}
\usepackage{amsmath}
\usepackage{amssymb}
\usepackage{stackrel}
\usepackage{graphicx}
\usepackage{epstopdf}



\begin{document}
\title{Quantum Zeno and anti-Zeno effects in the dynamics of non-degenerate
hyper-Raman processes coupled to two linear waveguides}



	\author{Moumita Das}
	\affiliation{Department of Physics, Malda College, Malda - 732101, India}	
		
	\author{Biswajit Sen}
	\email{bsen75@yahoo.co.in}
	\affiliation{Department of Physics, Vidyasagar Teachers' Training College, Midnapore - 721 101, India }
	
	\author{Kishore Thapliyal}
	\email{kishore.thapliyal@upol.cz}
	\affiliation{Palack\'{y} University Olomouc, Faculty of Science, Joint Laboratory of Optics of Palack\'{y} University and Institute of Physics of the Czech Academy of Sciences, 17. listopadu 1192/12, 779 00 Olomouc, Czech Republic}
	
	\author{Anirban Pathak}
	\email{anirban.pathak@jiit.ac.in}
	\affiliation{Jaypee Institute of Information Technology, A-10, Sector-62, Noida UP-201309, India}

\begin{abstract}
The effect of the presence of two probe waveguides on the dynamics
of hyper-Raman processes is studied in terms of quantum Zeno and anti-Zeno
effects. Specifically, the enhancement (diminution) of the evolution
of the hyper-Raman processes due to interaction with the probe waveguides
via evanescent waves is viewed as quantum Zeno (anti-Zeno) effect.
We considered the two probe waveguides interacting with only one of
the optical modes at a time. For instance, as a specific scenario,
it is considered that the two non-degenerate pump modes interact with
each probe waveguide linearly while Stokes and anti-Stokes modes do
not interact with the probes. Similarly, in another scenario, we assumed
both the probe waveguides interact with Stokes (anti-Stokes) mode
simultaneously. The present results show that quantum Zeno (anti-Zeno)
effect is associated with phase-matching (mismatching). However, we
did not find any relation between the presence of the quantum Zeno effect
and antibunching in the bosonic modes present in the hyper-Raman processes.
\end{abstract}

\maketitle

\section{Introduction}

As we are gradually approaching to obtain the benefits of the second
quantum revolution \cite{2QR} through the exciting developments
in the domain of quantum computing \cite{comp}, communication \cite{comm}
and sensing \cite{sens}, it is getting evident that we need to develop
efficient quantum control \cite{Qcon,Qcon2,Qcon3} as a fundamental
tool to realize the benefit of quantum technologies. In our endeavor
to achieve quantum control, quantum Zeno effect (QZE) \cite{misra1977zeno}
and quantum anti-Zeno effect (QAZE) may play a crucial role as QZE
(QAZE) speeds up (slows down) the time evolution of a quantum system
by frequent interaction or measurement \cite{misra1977zeno} and
thus allows us to control the evolution of the quantum system. The
phenomenon of QZE was introduced by Mishra and Sudarshan in 1977 \cite{misra1977zeno},
in analogy with one of the famous Zeno's paradoxes introduced by Greek
philosopher Zeno of Elea in the 5th century BC. Specifically, Mishra
and Sudarshan showed that if an unstable particle is continuously
measured, it will never decay. Noting the analogy of their observation
with the traditional Zeno's paradox, Mishra and Sudarshan referred
to this quantum phenomenon as \textit{Zeno's paradox in quantum theory}.
Though the term QZE owe its origin to the work of Mishra and Sudarshan,
much before their work Khalfin had studied nonexponential decay of
unstable atoms \cite{Khalfin} which was nothing but QZE (for a review
see Ref. \cite{Zeno-review-45}). 

Mishra and Sudarshan reported that the continuously measured unstable
particle will not decay and that would imply that the measurement
can slow down the evolution process. The existence of the opposite
phenomenon leading to speed up of the evolution of the quantum system
due to continuous measurement (or interaction) was soon predicted
theoretically and referred to as QAZE (see Refs. \cite{venugopalan2007zeno_review,facchi2001zeno-review,saverio-zeno-in-70-minutes,chi2-chi1-spie,PT,NZeno}).
Subsequent studies have revealed that QZE and QAZE can be witnessed
through a set of equivalent ways \cite{saverio-3-manifestation}.
In this work, we will use one of those ways that suits best for the
physical system of interest. Specifically, in what follows, QZE and
QAZE would be viewed as physical phenomena caused due to continuous
interaction between a system and one or more probes. In this continuous
interaction-type manifestation of QZE, we have considered a nonlinear
optical waveguide operating under a non-degenerate hyper-Raman process
involving two pump modes of different frequencies (referred hereafter
as system) is continuously interacting with two probe waveguides (probes).
The effect of these probe waveguides on the dynamics of the system
is to be studied and quantified in terms of increase/decrease in the
average boson numbers in all the modes not interacting directly with
the probes. For instance, when the probes interact with the pump modes
of the system (hyper-Raman active waveguide), the changes in the average
photon numbers of Stokes and anti-Stokes modes as well as the phonon
number are quantified. Similar manifestation of QZE was used earlier
to report QZE in Raman processes \cite{thun2002zeno-raman}. Later
on, some of the present authors reported QZE and/or QAZE in an asymmetric
and a symmetric nonlinear optical couplers \cite{chi2-chi1-spie,NZeno},
parity-time symmetric linear optical coupler \cite{PT} and hyper-Raman
waveguide interacting with a probe waveguide \cite{moumita-zeno-1}.
Extending the earlier works on QZE and QAZE in different nonlinear
optical systems \cite{chi2-chi1-spie,NZeno,thun2002zeno-raman,moumita-zeno-1},
here we aim to study a more general and mathematically complex system
where a hyper-Raman active waveguide is probed by two waveguides via
evanescent waves with a restriction that the two probe waveguides
interact with only one of the pump modes at a time. For instance,
the two non-degenerate pump modes interact with each probe waveguide
linearly while Stokes and anti-Stokes modes do not interact with the
probes. The beauty of the present work is that apart from rich dynamics
of the present system which is very general in nature, most of the
results reported earlier \cite{chi2-chi1-spie,NZeno,thun2002zeno-raman,moumita-zeno-1}
in the context of QZE and QAZE in Raman processes can be obtained
as special cases of the results obtained in the present work.

To perform the above aimed investigation, a specific perturbative
technique which is usually referred to as Sen-Mandal technique is
used \cite{kishore2014co-coupler,moumita-zeno-1,kishore2014contra,mandal2004co-coupler,bsen1}.
Specifically, this technique is used here to obtain analytic operator
solution of the Heisenberg's equation of motions corresponding the
physical system of interest (i.e., non-degenerate hyper-Raman processes
coupled to two linear waveguides). As these solutions are known to
be better than the solutions obtained by conventional short-length
approach \cite{kishore2014co-coupler,kishore2014contra,mandal2004co-coupler,bsen1}
and not restricted by the length, these are expected to yield more
accurate results if the spatial evolution of the field operators obtained
through this method is used here to study the dynamics of QZE and
QAZE processes in our system. The filed operator are also used to
study the possibility of single mode and intermodal antibunching with
an intention to reveal a relation between antibunching (a representative
nonclassical phenomenon) and QZE/QAZE. Though no such relation between
QZE/QAZE is revealed, antibunching is observed in some cases, and
it's found that the 

Till late nineties hardly any practical applications of QZE was proposed.
In fact, QZE remained as a topic of theoretical interest until it
was experimentally realized using a set of alternate approaches \cite{kwait-int.-free.measurement-expt.,experimental-zeno-1,experimental-zeno-2}.
Early experimental realizations of QZE has been followed by exciting
recent experiments having much more controls and precision \cite{experimental-zeno-2023}.
Experimental realization of QZE paved the way for various applications
of QZE \cite{kwait-int.-free.measurement-expt.,counterfactual-quantum-computation,Zubairy1,zeno-tomography-hradil,zeno-tomography2},
ranging from the enhancement of the resolution of absorption tomography
\cite{zeno-tomography-hradil,zeno-tomography2} to the demonstration
of entangling gate between two effectively non-interacting transmon
qubits \cite{zeno-gate}, the reduction of communication complexity
\cite{Q-com-comp} to the QZE and QAZE based noise spectroscopy \cite{noise-probe}
to the utilization of QZE to achieve improved precision of metrology
in presence of non-Markovian noise \cite{Long-non-Markovian-1}.
Interesting applications of QZE and QAZE are also found in opposing
decoherence by restricting the dynamics of the system in a decoherence-free
subspace \cite{Decoherence-free}, quantum interrogation measurement
\cite{kwait-int.-free.measurement-expt.}, counterfactual secure
quantum communication \cite{Zubairy1,coun-comEx}, isolating quantum
dot from its surrounding electron reservoir \cite{manipulation-of-Qdots},
protecting the entanglement between two interacting atoms \cite{protecting-entanglement}.
Even more practical problems like portfolio optimization problem is
proposed to be addressed using QZE \cite{portfolio-optmization},
and attempts have been made to study QZE in the macroscopic system,
like a large black hole \cite{Zeno-blackhole} and nonlinear waveguides
\cite{macroscopic-zenoPRL}, nonlinear optical couplers \cite{moumita-zeno-1}.
Present study is motivated by the great possibilities of application
of the QZE and QAZE established through these works, and the fact
that the physical system under consideration is extremely general
and experimentally realizable \cite{obrien1,obrien2}. 

The rest of the paper is organized as follows. In Section \ref{sec:Physical-system},
we have described the physical system of our interest (i.e., an optical
waveguide operating under a hyper-Raman process coupled with two linear
waveguides), and have presented the Heisenberg operator equations
of motion for the relevant operators. An explicit operator solution
of the Heisenberg operator equations is provided in Appendix \ref{sec:Appendix:-A},
but the solutions are used in Section \ref{sec:zeno-and-anti-zeno}
to study the dynamics of QZE and QAZE. Subsequently, in Section \ref{sec:zeno-and-anti-zeno},
possibilities of observing bunching and antibunching are discussed,
and the paper is finally concluded in Section \ref{sec:Conclusions}.

\begin{figure}
\begin{centering}
\includegraphics{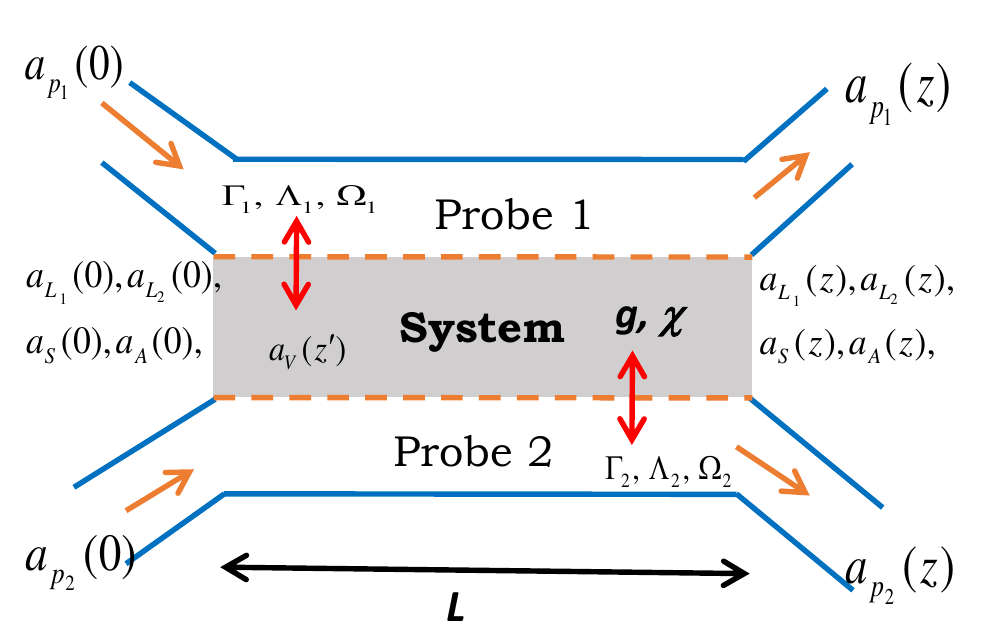}
\par\end{centering}
\caption{\label{fig:sys}(Color online) The schematic diagram of the two waveguides
(Probe 1 and Probe 2) interacting with a hyper-Raman active nonlinear
waveguide of length $L$ (system). All the parameters are defined
in the text. }

\end{figure}

\section{Physical system\label{sec:Physical-system}}

In our present study, we are considering a nonlinear waveguide of
length $L$ which produces non-degenerate hyper-Raman effect. The
system waveguide is coupled with two linear waveguides, and the coupling
has taken place in between the probe mode(s) and pump mode(s), Stokes
mode and anti-Stokes mode independently (cf. Fig. \ref{fig:sys}).
The complete system can be realised experimentally as the linear waveguides
are on either side of the system waveguide so that they exchange by
energy through evanescent waves. The presence of losses and the unavoidable
quantum noise arises from the system is negligible as we consider
one passage arrangement only and furthermore, it is expected that
the damping term reduce the quantum effects only. The effective momentum
operator involving the field modes which represents the physical system
of hyper-Raman processes is connected with two linear waveguides is
given by 
\begin{equation}
\begin{array}{lcl}
G & = & k_{p_{1}}a_{p_{1}}^{\dagger}a_{p_{1}}+k_{p_{2}}a_{p_{2}}^{\dagger}a_{p_{2}}+k_{L_{1}}a_{L_{1}}^{\dagger}a_{L_{1}}+k_{L_{2}}a_{L_{2}}^{\dagger}a_{L_{2}}+k_{S}a_{S}^{\dagger}a_{S}+k_{V}a_{V}^{\dagger}a_{V}+k_{A}a_{A}^{\dagger}a_{A}\\
 & + & \left(ga_{L_{1}}a_{L_{2}}a_{S}^{\dagger}a_{V}^{\dagger}+\chi a_{L_{1}}a_{L_{2}}a_{V}a_{A}^{\dagger}+{\rm H.c.}\right)+\left(\underset{l=1,2}{\sum}a_{p_{l}}^{\dagger}\left(\Gamma_{l}a_{L_{l}}+\Lambda_{l}a_{S}+\Omega_{l}a_{A}\right)+{\rm H.c.}\right),
\end{array}\label{eq:hamiltonian}
\end{equation}
where H.c. stands for the Hermitian conjugate. Throughout the present
paper, we use $\hbar=1$. The annihilation (creation) operators $a_{p_{i}}$(
$a_{p_{i}}^{\dagger})$ correspond to the probe mode in the two waveguides
which interact with the physical system undergoing a hyper-Raman process.
The operators $a_{L_{i}}(a_{L_{i}}^{\dagger}),\,a_{S}(a_{S}^{\dagger}),\,a_{V}\left(a_{V}^{\dagger}\right),\,a_{A}(a_{A}^{\dagger})$
in the momentum operator $G$ correspond to the two laser (pump) modes,
Stokes mode, vibration (phonon) mode, and anti-Stokes mode, respectively,
in the system. Here, $k_{p_{1}},$ $k_{p_{2}},$ $k_{L_{1}},$ $k_{L_{2}},$
$k_{S}$, $k_{V}$, and $k_{A}$ are the corresponding wavevectors
propagating along the $z$ axis. The parameters $g$ and $\chi$ denote
the Stokes and anti-Stokes coupling constants, respectively. $\Gamma_{l}$
denotes the interaction between the probe mode and pump mode and $\Lambda_{l}\,(\Omega_{l})$
is the interaction constant between the probe and the Stokes (anti-Stokes)
modes. In order to study the possibility of quantum Zeno (anti-Zeno)
effect as well as the quantum statistical properties of field modes
in the hyper-Raman processes coupled with two probe waveguides, we need
simultaneous solutions of the following Heisenberg operator equations
of motion for various field and phonon operators
\begin{equation}
\begin{array}{lcl}
\left.\overset{.}{a_{p_{j}}}\left(z\right)\right|_{j\in\{1,2\}} & = & i\left(k_{p_{j}}a_{p_{j}}+\Gamma_{j}a_{L_{j}}+\Lambda_{j}a_{S}+\Omega_{j}a_{A}\right),\\
\left.\overset{.}{a_{L_{i}}}\left(z\right)\right|_{i,\,j\in\{1,2\}} & = & i\left(k_{i}a_{L_{i}}+ga_{L_{j}}^{\dagger}a_{S}a_{V}+\chi a_{L_{j}}^{\dagger}a_{V}^{\dagger}a_{A}+\Gamma_{i}a_{p_{i}}\right),\\
\overset{.}{a_{S}}\left(z\right) & = & i\left(k_{S}a_{S}+ga_{L_{1}}a_{L_{2}}a_{V}^{\dagger}+\Lambda_{1}a_{p_{1}}+\Lambda_{2}a_{p_{2}}\right),\\
\overset{.}{a_{V}}\left(z\right) & = & i\left(k_{V}a_{V}+ga_{L_{1}}a_{L_{2}}a_{S}^{\dagger}+\chi a_{L_{1}}^{\dagger}a_{L_{2}}^{\dagger}a_{A}\right),\\
\overset{.}{a}_{A}\left(z\right) & = & i\left(k_{A}a_{A}+\chi a_{L_{1}}a_{L_{2}}a_{V}+\Omega_{1}a_{p_{1}}+\Omega_{2}a_{p_{2}}\right),
\end{array}\label{equationofmotion}
\end{equation}
where i,~$j\in\{1,2\}$ such that $i\neq j$ as well as $i$ and
$j$ satisfy modulo 2 algebra. The above set of coupled nonlinear
differential operator equations are not exactly solvable in the closed
analytical form under weak pump condition, but when the pump is very
strong the operator $a_{L_{i}}$ can be replaced by a $c$-number
and the above set of equations can be solved exactly. In order to
solve these equations under weak pump approximation, we use the perturbative
approach named as Sen-Mandal technique \cite{kishore2014co-coupler,moumita-zeno-1,kishore2014contra,mandal2004co-coupler,bsen1}
as the dimensionless quantities $gz,$ $\chi z,$$\Lambda_{l}z,$
$\Omega_{l}z$ and $\Gamma z$ are small compared to the unity. It
is already established that the solutions from this perturbative technique
are better than the well-known short-time approximation. The details
of the solution is explained in the Appendix \ref{sec:Appendix:-A}.
In what follows, this perturbative solutions are exploited to investigate
the Zeno dynamics as well as the quantum statistical properties of
the radiation field of various field modes.

\section{Quantum Zeno and anti-Zeno effect\label{sec:zeno-and-anti-zeno}}

In order to investigate the quantum Zeno and anti-Zeno effects in
hyper-Raman active medium coupled to two nonlinear waveguides, we
consider the initial composite coherent state arise from product of
the initial coherent states of probe $\left(\left|\alpha_{p_{i}}\right\rangle \right)$,
pump $\left(\left|\alpha_{L_{i}}\right\rangle \right)$, Stokes $\left(\left|\beta\right\rangle \right)$,
phonon $\left(\left|\gamma\right\rangle \right)$ and anti-Stokes
$\left(\left|\delta\right\rangle \right)$ modes, respectively. Hence,
the initial state is given by
\begin{equation}
\begin{array}{lcl}
\left|\psi(0)\right\rangle  & = & \left|\alpha_{p_{i}}\right\rangle \otimes\left|\alpha_{L_{i}}\right\rangle \otimes\left|\beta\right\rangle \otimes\left|\gamma\right\rangle \otimes\left|\delta\right\rangle ,\end{array}\label{initialstate}
\end{equation}
where $i\in\{1,2\}$ and the field operator $a_{p_{i}}$ operating
on the initial state gives 
\begin{equation}
\begin{array}{lcl}
a_{p_{i}}(0)\left|\psi(0)\right\rangle  & =\alpha_{p_{i}} & \left|\alpha_{p_{i}}\right\rangle \otimes\left|\alpha_{L_{i}}\right\rangle \otimes\left|\beta\right\rangle \otimes\left|\gamma\right\rangle \otimes\left|\delta\right\rangle ,\end{array}
\end{equation}
where $\alpha_{p_{i}}$ is the complex eigen value and the $\left|\alpha_{p_{i}}\right|^{2}$
is the average number of photons in the probe mode $a_{p_{i}}$. In
the similar manner, $\alpha_{i,}$$,$$\beta,$ $\gamma$ and $\delta$
are the complex amplitude corresponding to the pump, Stokes, phonon
and anti-Stokes modes, respectively. 

As noted in the previous section, QZE is referred to the slowing down
of quantum evolution due to continuous observation and the phenomenon
can be defined in many ways. In the present study, our system is coupled
to two probe waveguides and we want to observe how the presence of
interaction with the probe mode can affect the photon numbers in non-degenerate
hyper-Raman processes involving various field modes. Hence we may
define the Zeno parameter \cite{NZeno} as:
\begin{equation}
\begin{array}{lcl}
{\color{red}{\normalcolor Z_{j}^{L,S,A}}} & = & \left\langle N_{j}\right\rangle -\left\langle N_{j}\right\rangle _{\varGamma=\Omega=\Lambda=0},\end{array}\label{zeno-definition}
\end{equation}
where $j\in\{S,V,A\}$ which refers to Stokes, vibration phonon and
anti-Stokes, respectively, and $N_{j}$ refers to the number operator
of the $j^{th}$ mode. The superscripts $L,S,A$ correspond to Zeno
parameter of $j^{th}$ mode when the probe interacts with the Laser
(pump) mode, Stokes mode and anti-Stokes mode. The presence of QZE/QAZE
can be easily understood using the condition $Z_{j}<0\left(Z_{j}>0\right).$
It is clear from the equation (\ref{zeno-definition}) that the QZE
(QAZE) of a particular mode originates as a consequence of coupling
and would depend on the presence of the coupling constant $\Lambda_{l},\Omega_{l}$
and $\Gamma_{l}$ on the number operator of that mode. Now, using
the perturbative solution reported in the Appendix \ref{sec:Appendix:-A},
the Zeno parameter for these three modes are obtained as 
\begin{equation}
\begin{array}{lcl}
{\normalcolor {\color{red}{\normalcolor Z_{S}^{L,S,A}}}} & = & \left|l_{3}\right|^{2}\left|\alpha_{p_{1}}\right|^{2}+\left|l_{4}\right|^{2}\left|\alpha_{p_{2}}\right|^{2}+\left[\left\{ l_{1}l_{3}^{*}\alpha_{p_{1}}^{*}\beta+l_{1}l_{4}^{*}\alpha_{p_{2}}^{*}\beta+l_{1}l_{8}^{*}\alpha_{p_{1}}^{*}\alpha_{L_{2}}^{*}\beta{\normalcolor }\gamma\right.\right.\\
 & + & l_{1}l_{9}^{*}\alpha_{p_{2}}^{*}\alpha_{L_{1}}^{*}\beta\gamma+l_{1}l_{10}^{*}\alpha_{L_{1}}^{*}\beta+l_{1}l_{11}^{*}\alpha_{L_{2}}^{*}\beta+l_{1}l_{12}^{*}\beta\delta^{*}+l_{1}l_{13}^{*}\beta\delta^{*}+l_{1}l_{17}^{*}\left|\beta\right|^{2}\\
 & + & \left.\left.l_{1}l_{18}^{*}\left|\beta\right|^{2}+l_{2}l_{3}^{*}\alpha_{p_{1}}^{*}\alpha_{L_{1}}\alpha_{L_{2}}\gamma^{*}+l_{2}l_{4}^{*}\alpha_{p_{2}}^{*}\alpha_{L_{1}}\alpha_{L_{2}}\gamma^{*}+l_{3}l_{4}^{*}\alpha_{p_{1}}\alpha_{p_{2}}^{*}+\mathrm{c.c.}\right\} \right],
\end{array}
\end{equation}

\emph{
\begin{equation}
\begin{array}{lcl}
Z_{V}^{L,S,A} & = & \left[\left\{ m_{1}m_{7}^{*}\alpha_{p_{1}}^{*}\alpha_{L_{2}}^{*}\beta\gamma+m_{1}m_{8}^{\ast}\alpha_{p_{2}}^{*}\alpha_{L_{1}}^{*}\beta\gamma+m_{1}m_{9}^{*}\alpha_{p_{1}}\alpha_{L_{1}}^{*}\alpha_{L_{2}}^{*}\gamma\right.\right.\\
 & + & m_{1}m_{10}^{*}\alpha_{p_{2}}\alpha_{L_{1}}^{*}\alpha_{L_{2}}^{*}\gamma+m_{1}m_{11}^{*}\alpha_{p_{1}}\alpha_{L_{2}}\gamma\delta^{*}+m_{1}m_{12}^{*}\alpha_{p_{2}}\alpha_{L_{1}}\gamma\delta^{*}\\
 & + & \left.\left.m_{1}m_{13}^{*}\alpha_{p_{1}}^{*}\alpha_{L_{1}}\alpha_{L_{2}}\gamma+m_{1}m_{14}^{*}\alpha_{p_{2}}^{*}\alpha_{L_{1}}\alpha_{L_{2}}\gamma+\mathrm{c.c.}\right\} \right],
\end{array}
\end{equation}
 }and 
\begin{equation}
\begin{array}{lcl}
{\color{red}{\normalcolor Z_{A}^{L,S,A}}} & = & \left|n_{3}\right|^{2}\left|\alpha_{p_{1}}\right|^{2}+\left|n_{4}\right|^{2}\left|\alpha_{p_{2}}\right|^{2}+\left[\left\{ n_{1}n_{3}^{*}\alpha_{p_{1}}^{*}\delta+n_{1}n_{4}^{*}\alpha_{p_{2}}^{*}\delta+n_{1}n_{8}^{*}\alpha_{p_{1}}^{*}\alpha_{L_{2}}^{*}\gamma^{*}\delta\right.\right.\\
 & + & n_{1}n_{9}^{*}\alpha_{p_{2}}^{*}\alpha_{L_{1}}^{*}\gamma^{*}\delta+n_{1}n_{10}^{*}\alpha_{L_{1}}^{*}\delta+n_{1}n_{11}^{*}\alpha_{L_{2}}^{*}\delta+n_{1}n_{12}^{*}\beta^{*}\delta+n_{1}n_{13}^{*}\beta^{*}\delta\\
 & + & n_{1}n_{17}^{*}\left|\delta\right|^{2}+n_{1}n_{18}^{*}\left|\delta\right|^{2}+n_{2}n_{3}^{*}\alpha_{p_{1}}^{*}\alpha_{L_{1}}\alpha_{L_{2}}\gamma+n_{2}n_{4}^{*}\alpha_{p_{2}}^{*}\alpha_{L_{1}}\alpha_{L_{2}}\gamma\\
 & + & \left.\left.+n_{3}n_{4}^{*}\alpha_{p_{1}}\alpha_{p_{2}}^{*}+\mathrm{c.c.}\right\} \right].
\end{array}
\end{equation}

We have considered that the two probe modes interact with one of the
pump, Stokes, and anti-Stokes modes at one time. Finally, we also
discuss the scenario where a probe interacts with Stokes mode while
another probe interacts with anti-Stokes mode. To begin with, we discuss
the case when two probe modes are interacting with the two non-degenerate
pump modes in the system. 

\subsection{Case~I: Interaction of the probe with pump mode only}

We can obtain the case of interaction of the pump mode with the probe
as $\Gamma_{l}\neq0$ and $\Lambda_{l}=0=\Omega_{l}$. In this case,
Zeno parameter for the Stokes modes can be computed as 
\begin{equation}
\begin{array}{lcl}
Z_{S}^{L} & = & \stackrel[j=1]{2}{\sum}\mathcal{C}_{p_{j}}^{L}Z_{S_{j}}^{L},\end{array}\label{eq:ZSL}
\end{equation}
where 
\[
\begin{array}{lcl}
Z_{S_{j}}^{L} & = & \frac{1}{\Delta k_{S}\left(\Delta k_{S}+\Delta k_{L_{j}}\right)\Delta k_{L_{j}}}\left[\frac{\Delta k_{S}}{2}\left\{ \cos\left(\Delta\theta_{S}+\Delta\phi_{L_{j}}+\Delta k_{S}z+\Delta k_{L_{j}}z\right)-\cos\left(\Delta\theta_{S}+\Delta\phi_{L_{j}}+\Delta k_{S}z\right)\right\} \right.\\
 & - & \left.\frac{\Delta k_{L_{j}}}{2}\left\{ \cos\left(\Delta\theta_{S}+\Delta\phi_{L_{j}}+\Delta k_{S}z\right)-\cos\left(\Delta\theta_{S}+\Delta\phi_{L_{j}}\right)\right\} \right],
\end{array}
\]
and $\mathcal{C}_{p_{j}}^{L}=4g\varGamma_{j}\left|\alpha_{p_{j}}\right|\left|\alpha_{L_{j\oplus1}}\right|\left|\beta\right|\left|\gamma\right|$
with the phase difference parameters in Stokes process $\Delta\theta_{S}=(\phi_{S}+\phi_{V}-\sum_{j}\phi_{L_{j}})$,
anti-Stokes process $\Delta\theta_{A}=(\sum_{j}\phi_{L_{j}}+\phi_{V}-\phi_{A})$,
and laser-probe interaction $\Delta\phi_{L_{j}}=(\phi_{L_{j}}-\phi_{p_{j}})$.
We have also written the complex amplitudes of the initial coherent
states here in polar form as $\alpha_{m}=|\alpha_{m}|\exp[i\phi_{m}]:m\in\{p_{1},p_{2},L_{1},L_{2}\}$,
$\beta=|\beta|\exp[i\phi_{S}]$, $\gamma=|\gamma|\exp[i\phi_{V}]$,
and $\delta=|\delta|\exp[i\phi_{A}]$. Here, the phase mismatch parameters
$\Delta k_{S}=(k_{S}+k_{V}-k_{L_{1}}-k_{L_{2}})$ and $\Delta k_{A}=(k_{L_{1}}+k_{L_{2}}+k_{V}-k_{A})$
with $\Delta k_{L_{j}}=(k_{L_{j}}-k_{p_{j}})$. Here, $S$ in the
subscript corresponds to Zeno parameter of Stokes mode when the probe
is interacting with the laser mode, represented by superscript $L$.
As we study the QZE and QAZE in hyper-Raman active waveguide with
two probes, which forms a combined system that has a line of symmetry
along the length of the system waveguide, this results in resultant
quantity given in Eq. (\ref{eq:ZSL}) as sum of two symmetric terms
depending upon each probe-pump mode interaction independently. Further,
note that $\mathcal{C}_{p_{j}}^{L}$ is a positive phase independent
quantity (assuming the Stokes and anti-Stokes coupling constants as
real) and cannot control the transition between the quantum Zeno and
anti-Zeno effects. 

Similarly, we obtained the Zeno parameter for anti-Stokes mode in
this case as 
\begin{equation}
\begin{array}{lcl}
Z_{A}^{L} & = & \stackrel[j=1]{2}{\sum}\mathcal{D}_{p_{j}}^{L}Z_{A_{j}}^{L},\end{array}\label{eq:ZAL}
\end{equation}
where
\[
\begin{array}{lcl}
Z_{A_{j}}^{L} & = & \stackrel[j=1]{2}{\sum}\frac{1}{\Delta k_{A}\left(\Delta k_{A}-\Delta k_{L_{j}}\right)\Delta k_{L_{j}}}\left[-\frac{\Delta k_{A}}{2}\left\{ \cos\left(\Delta\theta_{A}-\Delta\phi_{L_{j}}+\Delta k_{A}z-\Delta k_{L_{j}}z\right)-\cos\left(\Delta\theta_{A}-\Delta\phi_{L_{j}}+\Delta k_{A}z\right)\right\} \right.\\
 & - & \left.\frac{\Delta k_{L_{j}}}{2}\left\{ \cos\left(\Delta\theta_{A}-\Delta\phi_{L_{j}}+\Delta k_{A}z\right)-\cos\left(\Delta\theta_{A}-\Delta\phi_{L_{j}}\right)\right\} \right],
\end{array}
\]
and $\mathcal{D}_{p_{j}}^{L}=\mathcal{C}_{p_{j}}^{L}\frac{\chi}{g}\frac{\left|\delta\right|}{\left|\beta\right|}$
is a positive quantity deciding the depth of Zeno parameter. We further
obtain that the Zeno parameter for phonon mode is

\begin{equation}
\begin{array}{lcl}
Z_{V}^{L} & = & Z_{S}^{L}-Z_{A}^{L}.\end{array}\label{eq:ZLV}
\end{equation}
This relation can be viewed as a consequence of the constant of motion
$\frac{d}{dt}\left(N_{S}\left(t\right)-N_{A}\left(t\right)-N_{V}\left(t\right)\right)=0$
associated with the momentum operator.

From the values of $\mathcal{C}_{p_{j}}^{L}$ and $\mathcal{D}_{p_{j}}^{L}$,
we can observe that all Zeno parameter $Z_{k}^{L}=0\forall\,k\in\left\{ S,A,V\right\} $
in the spontaneous case when Stokes, anti-Stokes, and phonon modes
are initially in vacuum, i.e., $\beta=0=\gamma=\delta$. Even, in
the partially stimulated case when one or more (but not all) modes
are initially in the vacuum (here we consider $\gamma=0$ and $\beta\neq0\neq\delta$
), all the Zeno parameters remain zero as $\mathcal{C}_{p_{j}}^{L}=0=\mathcal{D}_{p_{j}}^{L}$
and the same is observed when $\gamma\neq0$, and $\beta=\delta=0.$
Therefore, we obtain non-trivial cases for Stokes mode when $\gamma\neq0\neq\beta$,
anti-Stokes mode when $\gamma\neq0\neq\delta$, and photon mode in
both these cases. Therefore, in what follows, we discuss the stimulated
case when none of the mode is initially in the vacuum. 

\begin{figure}
\centering{}\includegraphics[scale=0.3]{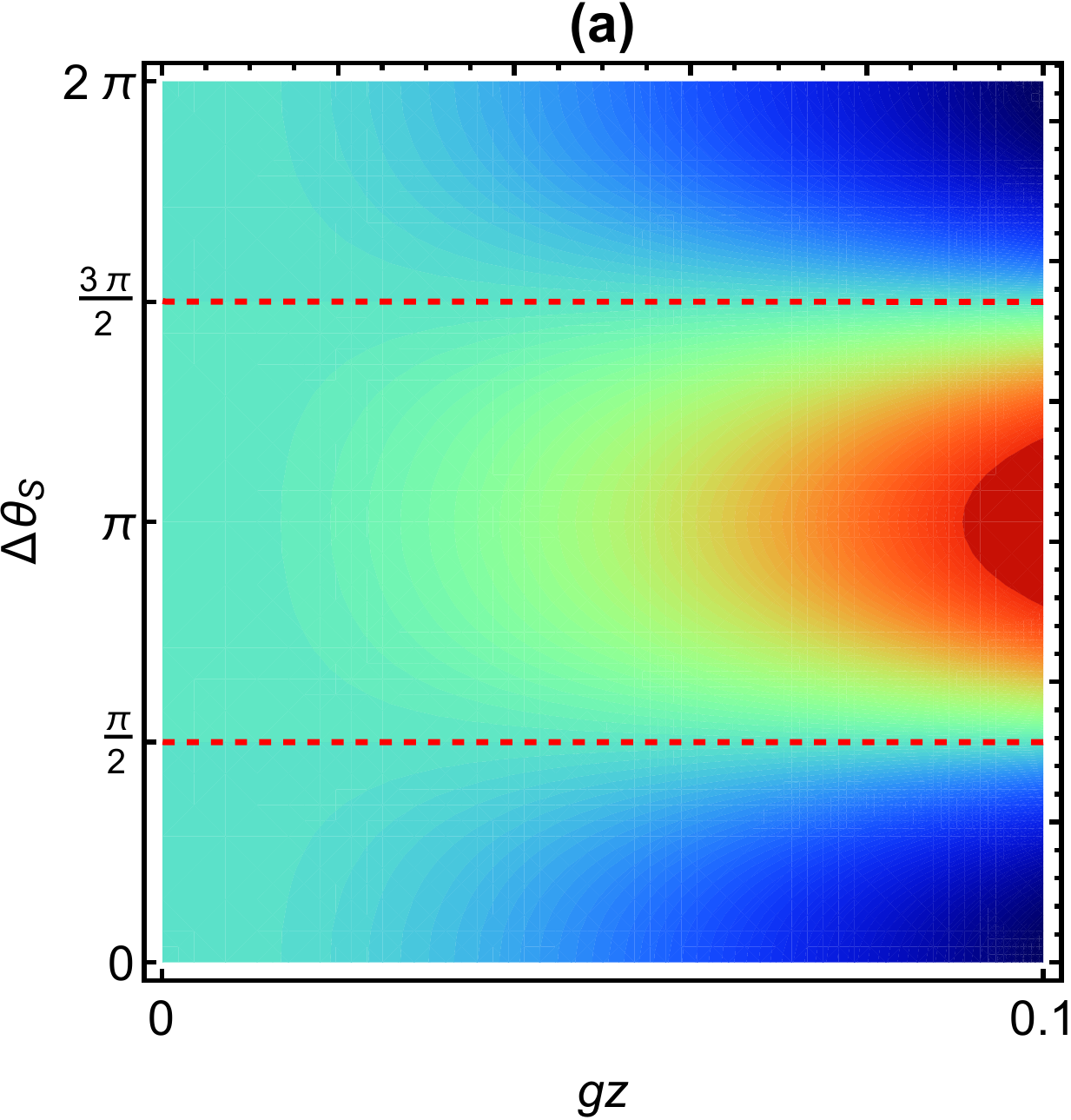} \includegraphics[scale=0.4]{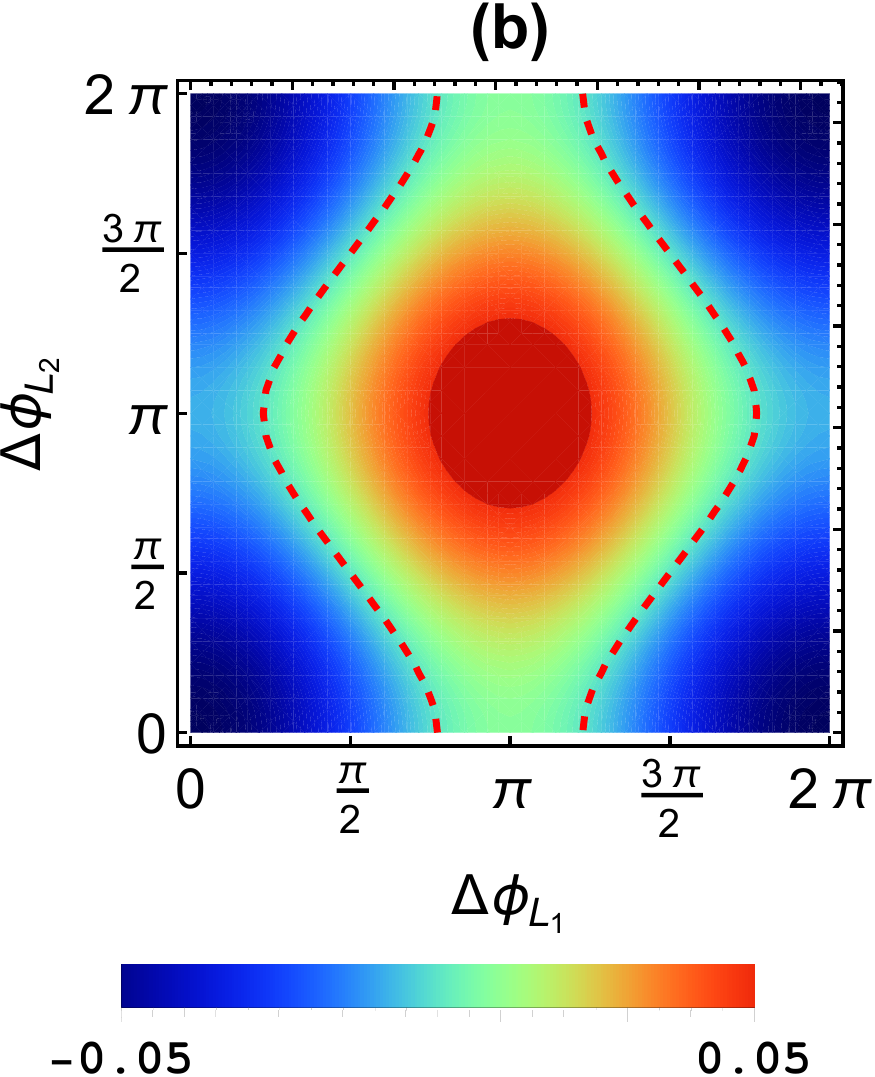}
\includegraphics[scale=0.3]{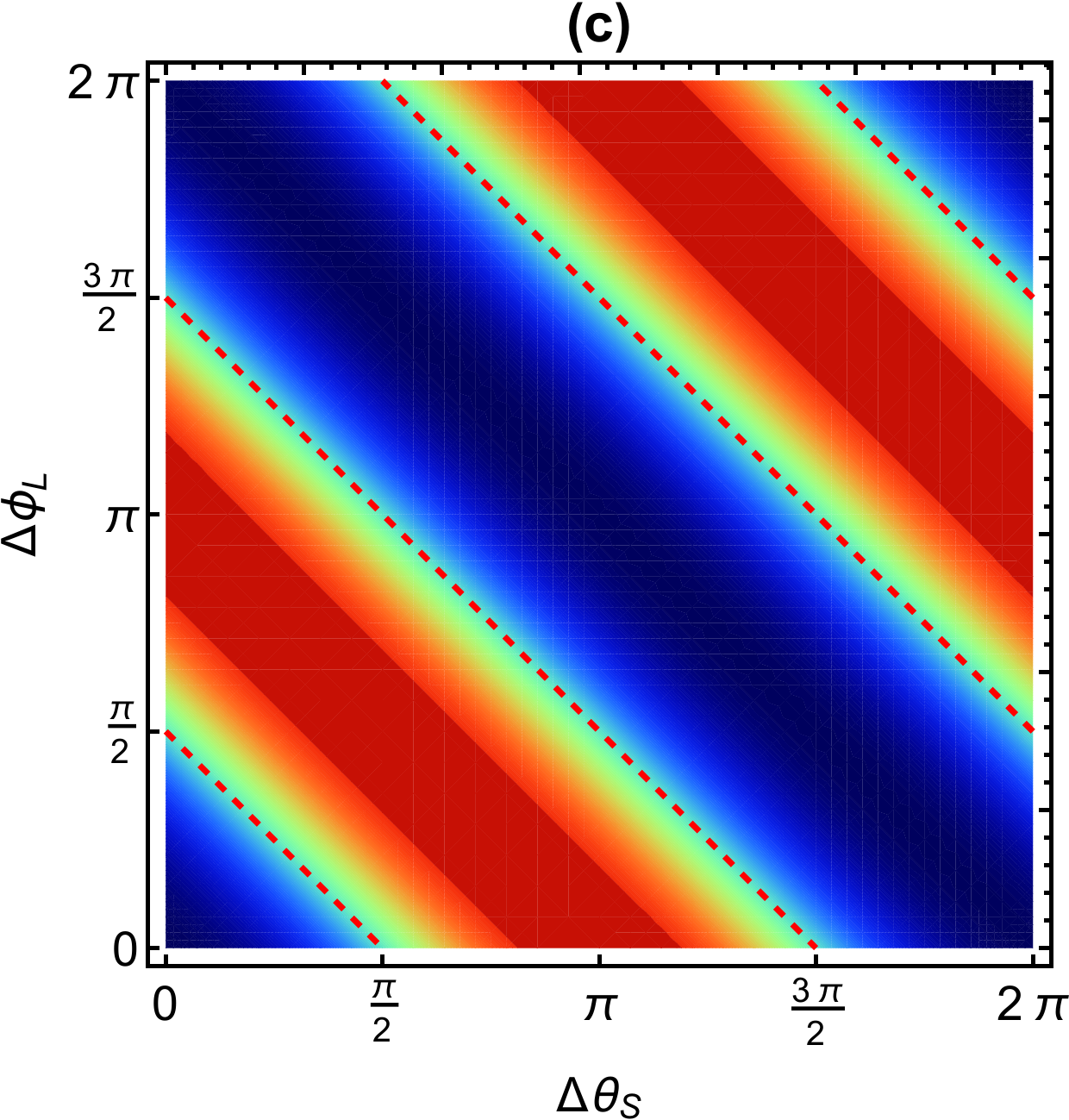}\caption{\label{fig:ZLS}(Color online) Spatial evolution of Zeno parameter
with phase matching $\mathcal{Z}_{S}^{L}$ for Stokes mode with (a)-(c)
phase difference parameters in Stokes $\Delta\theta_{S}$ and probe-system
$\Delta\phi_{L_{j}}$ interactions with $\Delta\theta_{S}=0=\Delta\phi_{L_{j}}$
and $gz=0.1$ wherever needed. In (c), $\Delta\phi_{L_{1}}=\Delta\phi_{L_{2}}=\Delta\phi_{L}$.
We have also used $\chi=1.1g$, $\Gamma_{1}=\Gamma_{2}=1.15g$, $\alpha_{p_{1}}=5,\alpha_{p_{2}}=4,\alpha_{L_{1}}=8,\alpha_{L_{2}}=8.5,$
$\beta=7,$ $\gamma=0.01$, and $\delta=1$ here and in the rest of
the figures. }
\end{figure}
\begin{figure}
\centering{}\includegraphics[scale=0.3]{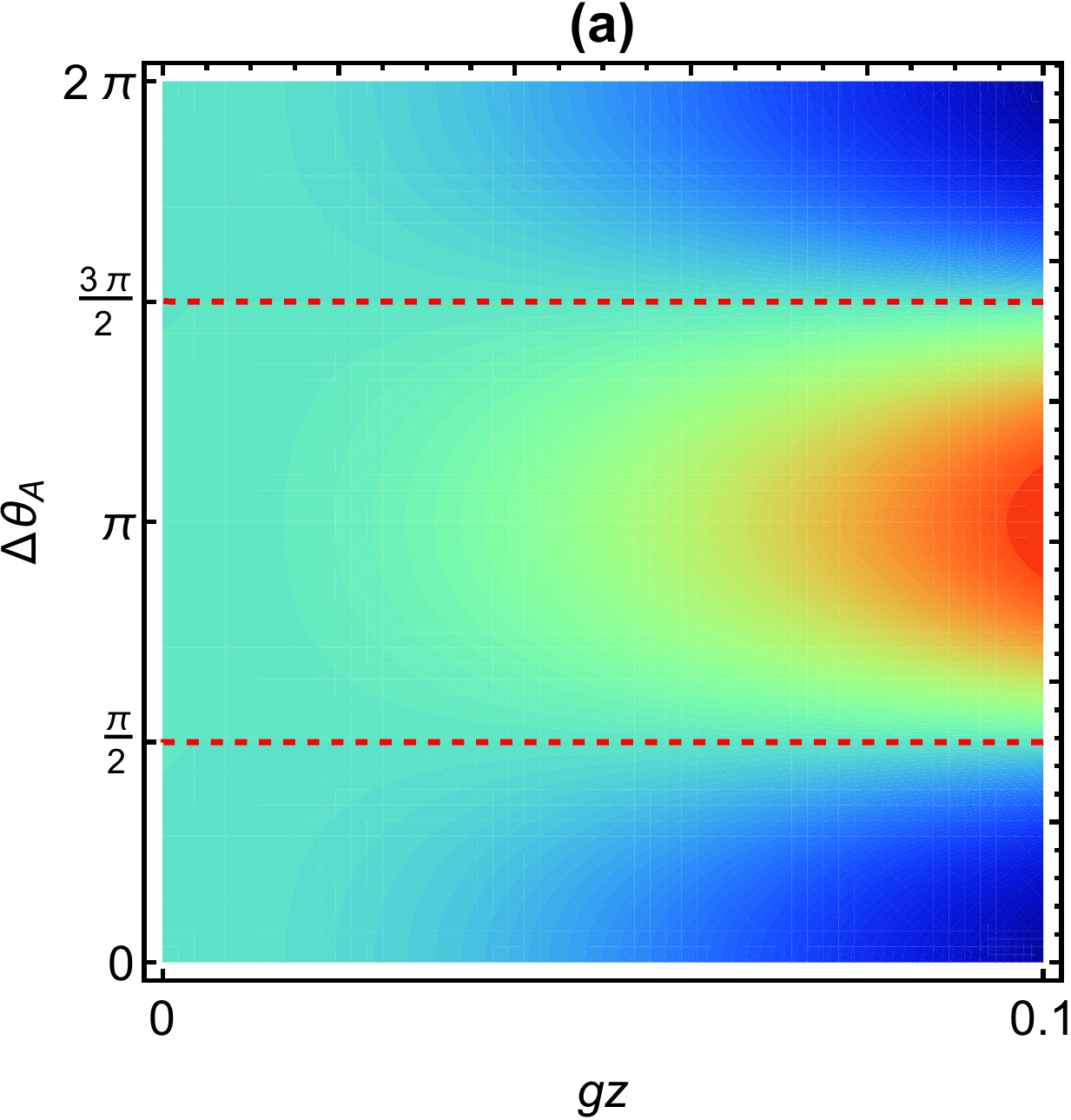} \includegraphics[scale=0.4]{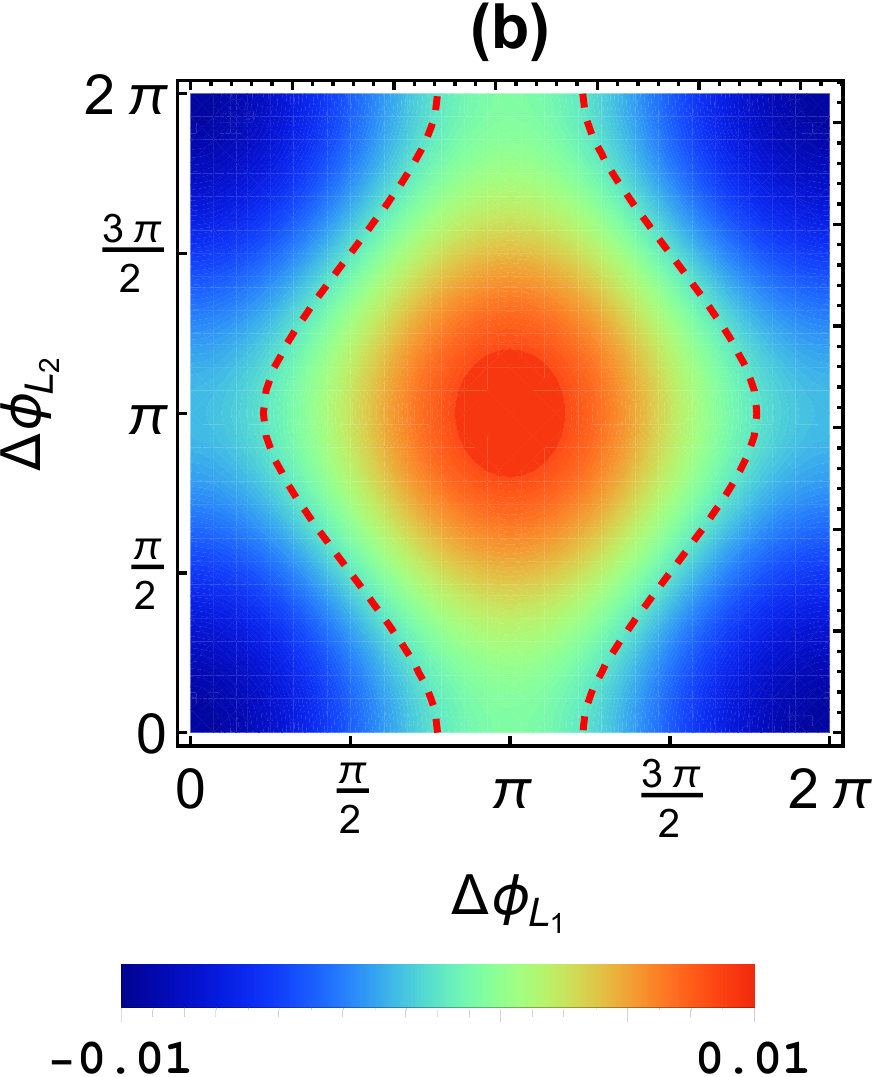}
\includegraphics[scale=0.3]{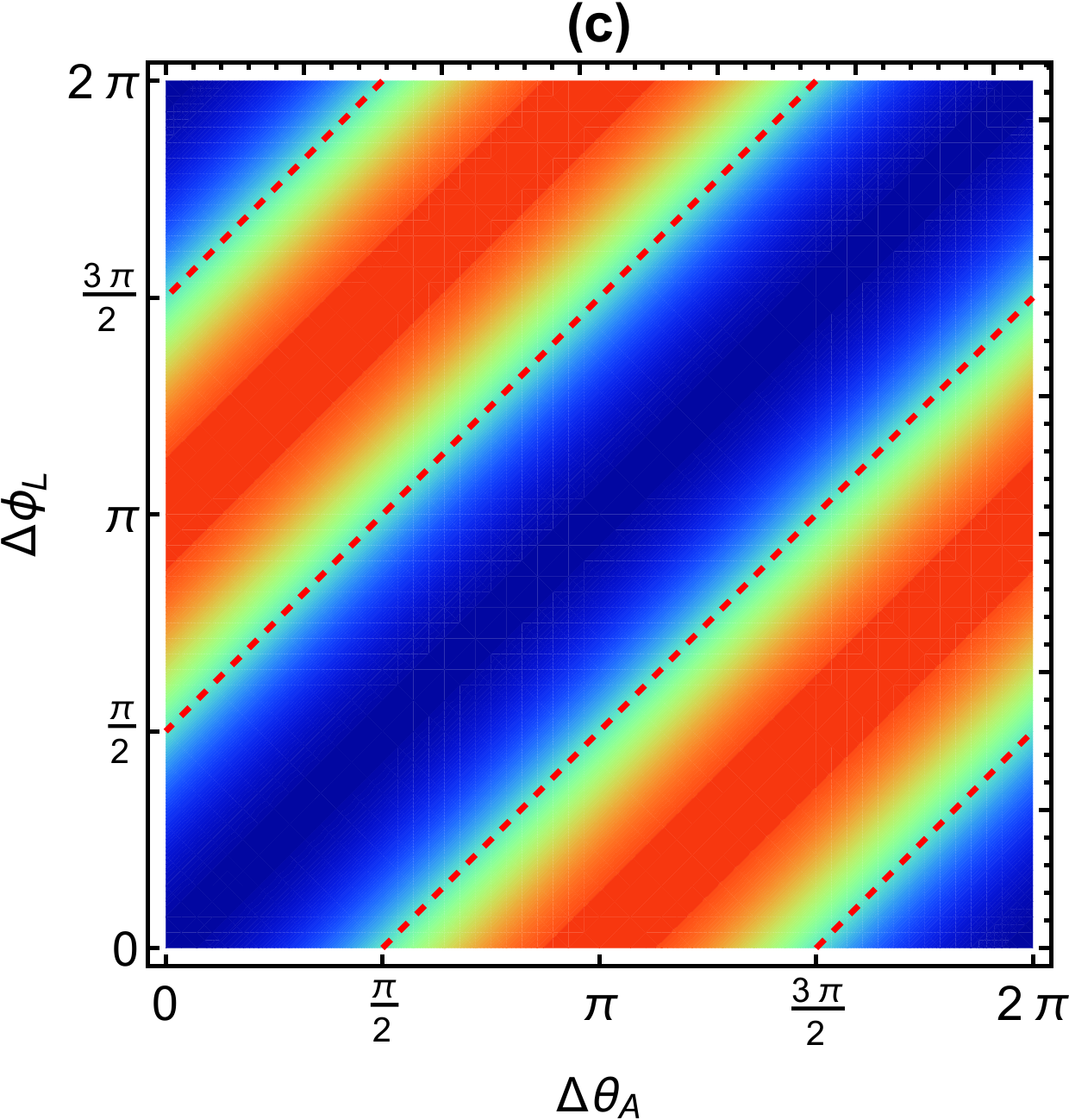}\caption{\label{fig:ZLA}(Color online) Spatial evolution of Zeno parameter
with phase matching $\mathcal{Z}_{A}^{L}$ for anti-Stokes mode with
(a)-(c) phase difference parameters in anti-Stokes $\Delta\theta_{A}$
and probe-system $\Delta\phi_{L_{j}}$ interactions with $\Delta\theta_{A}=0=\Delta\phi_{L_{j}}$
and $gz=0.1$ wherever needed. In (c), $\Delta\phi_{L_{1}}=\Delta\phi_{L_{2}}=\Delta\phi_{L}$. }
\end{figure}
An interesting scenario is obtained when the phase mismatch parameters
for Stokes, anti-Stokes, and pump-probe interaction $\Delta k_{S}z$,
$\Delta k_{A}z$, and $\Delta k_{L_{j}}z$, respectively, vanish.
In this case, we obtain the Zeno parameter in this case as
\[
\begin{array}{lcl}
\mathcal{Z}_{k}^{L} & = & \underset{\Delta k_{S}\rightarrow0,\Delta k_{A}\rightarrow0,\Delta k_{L_{j}}\rightarrow0}{\lim}Z_{k}^{L}\,\forall\,k\in\left\{ S,A,V\right\} \end{array}.
\]
Specifically, we have obtained for the Stokes and anti-Stokes modes
$\mathcal{Z}_{S}^{L}=-\frac{1}{4}\stackrel[j=1]{2}{\sum}\mathcal{C}_{p_{j}}^{L}z^{2}\cos\left(\Delta\theta_{S}+\Delta\phi_{L_{j}}\right)$
and $\mathcal{Z}_{A}^{L}=-\frac{1}{4}\stackrel[j=1]{2}{\sum}\mathcal{D}_{p_{j}}^{L}z^{2}\cos\left(\Delta\theta_{A}-\Delta\phi_{L_{j}}\right)$,
respectively. Further, we can obtain corresponding case for phonon
mode as $\mathcal{Z}_{V}^{L}=\mathcal{Z}_{S}^{L}-\mathcal{Z}_{A}^{L}$.
Thus, we can observe QZE in Stokes and anti-Stokes mode for $\Delta\theta_{S},\Delta\theta_{A}\in\left[-\frac{\pi}{2},\frac{\pi}{2}\right]$
if $\Delta\phi_{L_{j}}=0$ (as shown in Figs. \ref{fig:ZLS} (a) and
\ref{fig:ZLA} (a)). Further, notice that the dependence of the Zeno
parameter on $\Delta\phi_{L_{1}}=\Delta\phi_{L_{2}}=\Delta\phi_{L}$
is analogous to that of the Stokes and anti-Stokes phase difference
parameters if we consider $\Delta\theta_{S}=0=\Delta\theta_{A}$.
Therefore, we consider the case when $\Delta\phi_{L_{1}}\neq\Delta\phi_{L_{2}}$
and $\Delta\theta_{S}=0=\Delta\theta_{A}$ and observe that QAZE is
dominant when the phase difference is $\Delta\phi_{L_{1}}=\Delta\phi_{L_{2}}=\pi$
(cf. Figs. \ref{fig:ZLS} (b) and \ref{fig:ZLA} (b)). Variation of
$\Delta\phi_{L_{1}}=\Delta\phi_{L_{2}}=\Delta\phi_{L}$ and $\Delta\theta_{S}$
or $\Delta\theta_{A}$ shows that QZE is associated with phase matching
in the nonlinear process in the system as well as between probe and
the system (cf. Figs. \ref{fig:ZLS} (c) and \ref{fig:ZLA} (c)).
Further, we can observe that the variation in the Zeno parameter for
the Stokes mode is more prominent than that in anti-Stokes mode (cf.
Figs. \ref{fig:ZLS}-\ref{fig:ZLA}) due to $\mathcal{C}_{p_{j}}^{L}>\mathcal{D}_{p_{j}}^{L}$
as we have used $\frac{\chi}{g}\frac{\left|\delta\right|}{\left|\beta\right|}<1$,
the Zeno parameter of the phonon mode shows a similar behavior as
that of Stokes mode. However, if one uses $\chi\left|\delta\right|>g\left|\beta\right|$,
the QZE (QAZE) for phonon mode will be observed corresponding to the
QZE (QAZE) for the Stokes/anti-Stokes mode.

An interesting case of phase mismatch is $\Delta k_{S}z=-\Delta k_{A}z=\Delta k_{L_{j}}z$,
which supports the phonon mode as $2k_{V}=k_{A}-k_{S}$. In this case,
we obtain that the Zeno parameters can be obtained as $Z_{S}^{\prime L}=-\frac{1}{4}\stackrel[j=1]{2}{\sum}\mathcal{C}_{p_{j}}^{L}z^{2}\cos\left(\Delta k_{S}z+\Delta\theta_{S}+\Delta\phi_{L_{j}}\right)\mathrm{sinc}^{2}\left(\frac{\Delta k_{S}z}{2}\right)$
and $Z_{A}^{\prime L}=-\frac{1}{4}\stackrel[j=1]{2}{\sum}\mathcal{D}_{p_{j}}^{L}z^{2}\cos\left(\Delta k_{S}z-\Delta\theta_{A}+\Delta\phi_{L_{j}}\right)\mathrm{sinc}^{2}\left(\frac{\Delta k_{S}z}{2}\right)$.
After comparing these expressions with $\mathcal{Z}_{j}^{L}$ we can
observe that the phase mismatch modulates the Zeno parameter function
obtained for phase matching condition. We can observe a transition
between QZE and QAZE for $\Delta k_{S}z=\frac{\left(2n+1\right)\pi}{2}$
and $\Delta k_{S}z=2m\pi$ with integer $n$ and non-zero integer
$m$ if the phase difference parameters $\Delta\theta_{S}=\Delta\theta_{A}=0=\Delta\phi_{L_{j}}$. 

\begin{figure}
\centering{}\includegraphics[scale=0.6]{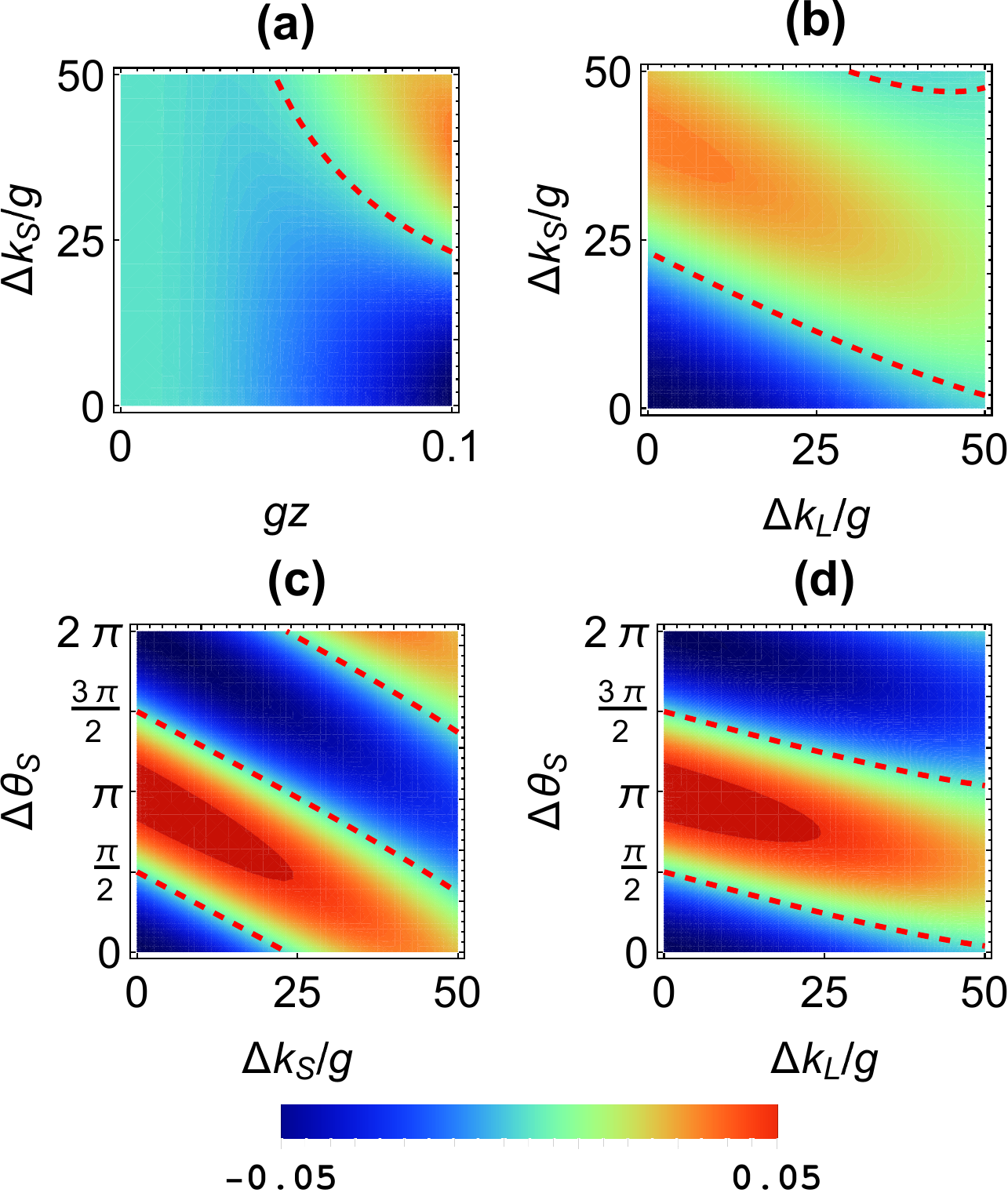}\caption{\label{fig:ZLS-det}(Color online) Spatial evolution of Zeno parameter
$Z_{S}^{L}$ for Stokes mode with (a)-(c) phase difference parameters
in Stokes $\Delta\theta_{S}$ and probe-system $\Delta\phi_{L_{j}}$
interactions as well as phase mismatches $\Delta k_{S}/g$ and $\Delta k_{L_{j}}/g=\Delta k_{L}/g\,\forall j\in\left\{ 1,2\right\} $.
We used $\Delta\theta_{S}=0=\Delta\phi_{L_{j}}$ and $gz=0.1$ wherever
needed.}
\end{figure}

If we consider the effect of phase mismatch $\Delta k_{S}z$ on the
Zeno parameter of the Stokes mode in the absence of any phase difference
$\Delta\theta_{S}=0=\Delta\phi_{L_{j}}$ as well as phase matching
in probe-pump interaction $\Delta k_{L_{j}}z=0$, we can observe that
a transition between QZE and QAZE is observed at $\approx0.742\pi$
(cf. Fig. \ref{fig:ZLS-det} (a)) unlike $0.5\pi$ in special case
of $Z_{S}^{\prime L}$. However, increasing the value of phase mismatch
in the probe-pump interaction $\Delta k_{L_{1}}z=\Delta k_{L_{2}}z=\Delta k_{L}z$,
we can attain this transition for smaller values of $\Delta k_{S}z$
(cf. Fig. \ref{fig:ZLS-det} (b)). Considering non-zero initial phase
difference in the Stokes interaction $\Delta\theta_{S}$ we can observe
the equivalence in the role of phase difference and phase mismatch
parameters (see Fig. \ref{fig:ZLS-det} (c)-(d)). Notice that change
in Stokes phase mismatch parameter is more sensitive parameter in
comparison to the phase mismatch in pump-probe interaction. 

\begin{figure}
\centering{}\includegraphics[scale=0.6]{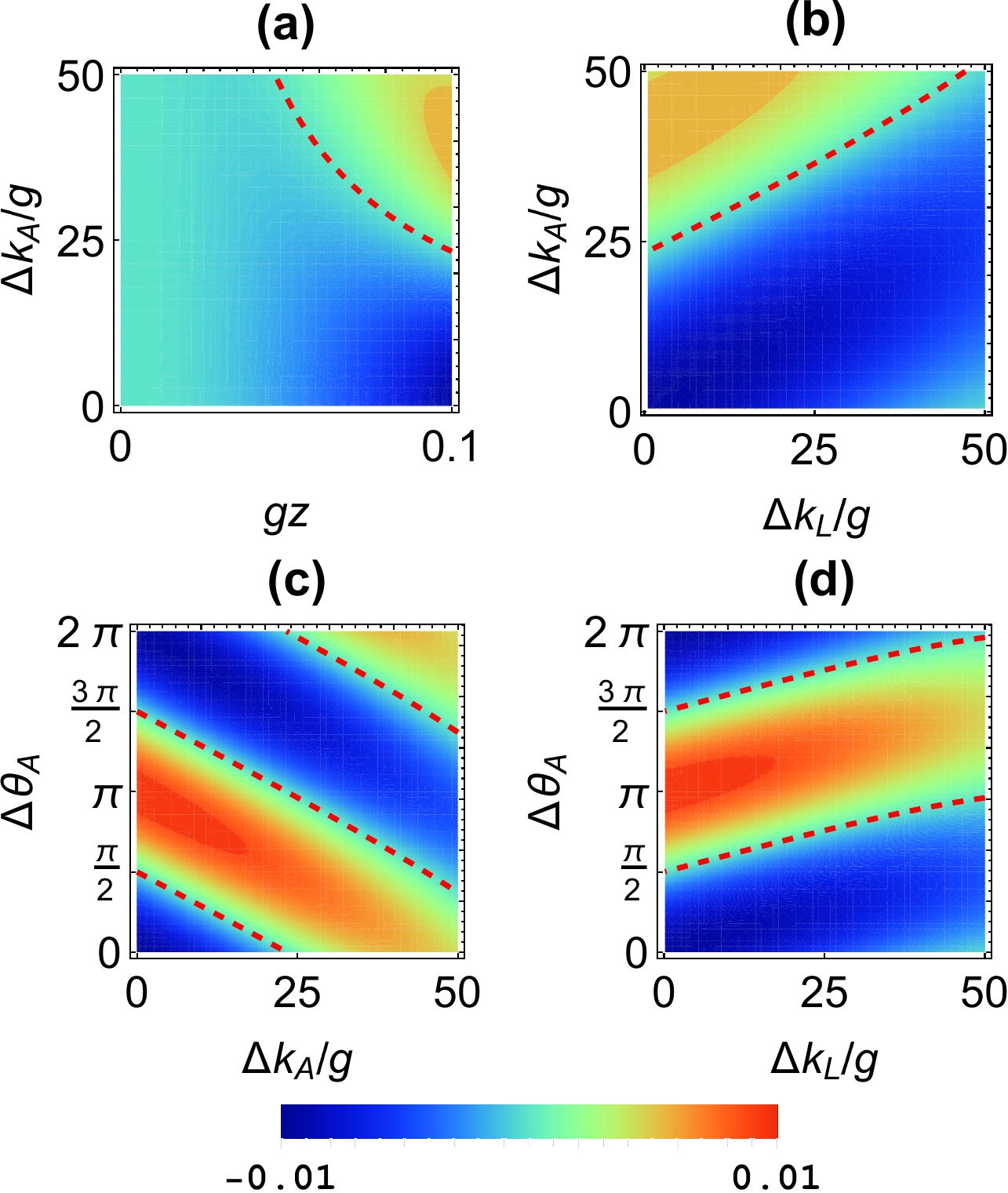}\caption{\label{fig:ZLA-det}(Color online) Spatial evolution of Zeno parameter
$Z_{A}^{L}$ for anti-Stokes mode with (a)-(d) phase difference parameters
in anti-Stokes $\Delta\theta_{A}$ and probe-system $\Delta\phi_{L_{j}}$
interactions as well as phase mismatches $\Delta k_{A}/g$ and $\Delta k_{L_{j}}/g=\Delta k_{L}/g\,\forall j\in\left\{ 1,2\right\} $.
We used $\Delta\theta_{A}=0=\Delta\phi_{L_{j}}$ and $gz=0.1$ wherever
needed.}
\end{figure}

The Zeno parameter of the anti-Stokes mode shows a similar dependence
on phase mismatch $\Delta k_{A}z$ as the Zeno parameter of Stokes
mode on the phase mismatch $\Delta k_{S}z$ in the absence of any
phase difference $\Delta\theta_{A}=0=\Delta\phi_{L_{j}}$ as well
as phase matching in probe-pump interaction $\Delta k_{L_{j}}z=0$
(cf. Figs. \ref{fig:ZLS-det} (a) and \ref{fig:ZLA-det} (a)). However,
increasing the value of phase mismatch in the probe-pump interaction
$\Delta k_{L_{1}}z=\Delta k_{L_{2}}z=\Delta k_{L}z$ enhance the values
of $\Delta k_{A}z$ required for a transition between quantum Zeno
and anti-Zeno effects (cf. Fig. \ref{fig:ZLA-det} (b)). Considering
non-zero initial phase difference in the anti-Stokes interaction $\Delta\theta_{A}$
we can observe that both phase difference $\Delta\theta_{A}$ and
phase mismatch $\Delta k_{A}z$ parameters are in phase and supports
each other (see Fig. \ref{fig:ZLA-det} (c)). While the phase mismatch
parameter $\Delta k_{L}z$ works out of phase with respect to the
phase difference parameter $\Delta\theta_{A}$. 

\begin{figure}
\centering{}\includegraphics[scale=0.6]{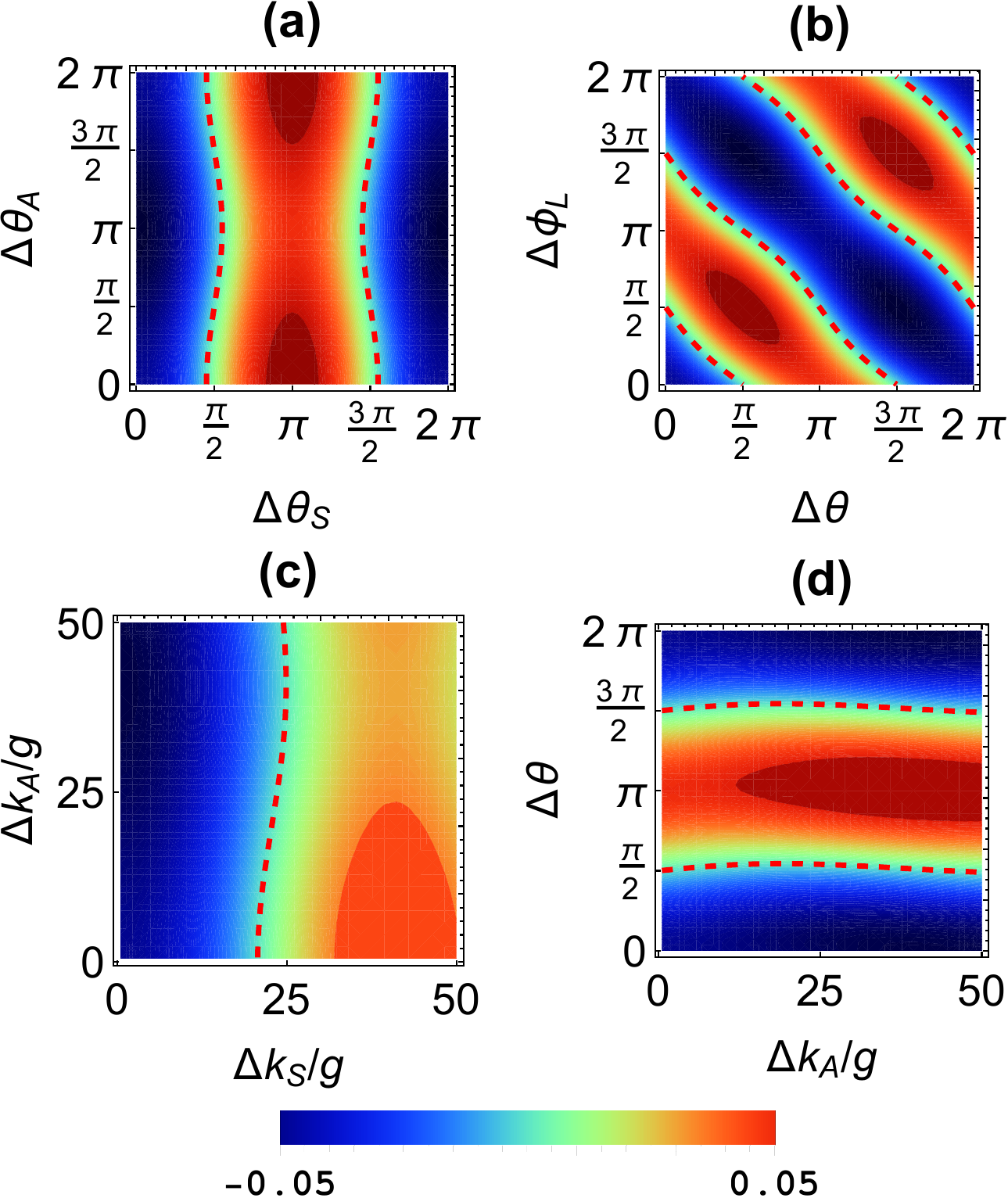}\caption{\label{fig:ZLV}(Color online) Spatial evolution of Zeno parameter
$Z_{V}^{L}$ for photon mode with (a)-(d) phase difference parameters
in Stokes $\Delta\theta_{S}$, anti-Stokes $\Delta\theta_{A}$ and
probe-system $\Delta\phi_{L_{j}}$ interactions as well as phase mismatches
$\Delta k_{S}/g$ and $\Delta k_{A}/g$. We used $\Delta\theta_{S}=\Delta\theta_{A}=\Delta\theta$
and $\Delta\phi_{L_{1}}=\Delta\phi_{L_{2}}=\Delta\phi_{L}$. The dependence
of $Z_{V}^{L}$ on the rest of the parameters is the similar as $Z_{S}^{L}$. }
\end{figure}

Under the usual condition $\chi\left|\delta\right|<g\left|\beta\right|$,
we observed that the QZE (QAZE) for phonon mode has the analogous
dependence to that of the QZE (QAZE) for the Stokes mode. For instance,
the dependence of Zeno parameter for phonon mode on the phase difference
parameters (assuming $\Delta\theta_{S}=\Delta\theta_{A}=\Delta\theta$)
is similar to that of Stokes/anti-Stokes modes (as shown in Figs.
\ref{fig:ZLS} (a)-(b) and \ref{fig:ZLA} (a)-(b)). However, variation
of both phase difference parameters of Stokes and anti-Stokes processes
reveals that QZE is associated with phase matching (see Fig. \ref{fig:ZLV}
(a)). We can also observe that due to the competing effects of phase
difference in Stokes and anti-Stokes processes (as shown in Figs.
\ref{fig:ZLS} (c) and \ref{fig:ZLA} (c)), we observe the combined
effect dominated by Stokes phase difference parameter and some modulations
due to anti-Stokes phase difference parameter (see Fig. \ref{fig:ZLV}
(b)). An increase in the phase mismatch parameter in Stokes (anti-Stokes)
process favors QAZE (QZE) (cf. Fig. \ref{fig:ZLV} (c)). However,
a phase change induced by the phase difference parameter $\Delta\theta=\pi$
makes anti-Stokes phase mismatch parameter to support QAZE (see Fig.
\ref{fig:ZLV} (d)).

\subsection{Case~II: Interaction of the probe with Stokes mode only }

Interaction of the Stokes mode with the probe can be obtained as $\Lambda_{l}\neq0$
and $\Gamma_{l}=0=\Omega_{l}$. We study the quantum Zeno and anti-Zeno
effects in this case for anti-Stoke and phonon modes only as Stokes
mode is interacting with the probe. In this case, Zeno parameter for
the anti-Stokes modes vanishes in the range of the validity of the
solution, i.e., $\begin{array}{lcl}
Z_{A}^{S} & = & 0\end{array}$. Here, $A$ in the subscript corresponds to Zeno parameter of anti-Stokes
mode when the probe is interacting with the Stokes mode, represented
by superscript $S$. It is worth mentioning here that the corresponding
parameter with short-length approximation technique is also fourth
order in rescaled length \cite{thun2002zeno-raman}. The Zeno parameter
for phonon mode in this case is 

\subsection*{{\normalsize{}
\begin{equation}
\protect\begin{array}{lcl}
Z_{V}^{S} & = & \stackrel[j=1]{2}{\sum}\frac{\mathcal{C}_{p_{j}}^{S}}{\Delta k_{S}\left(\Delta k_{S}-\Delta k_{S_{j}}\right)\Delta k_{S_{j}}}\left[\frac{\Delta k_{S}}{2}\left\{ \cos\left(\Delta\theta_{S}-\Delta\phi_{S_{j}}+\Delta k_{S}z-\Delta k_{S_{j}}z\right)-\cos\left(\Delta\theta_{S}-\Delta\phi_{S_{j}}+\Delta k_{S}z\right)\right\} \right.\protect\\
 & - & \left.\frac{\Delta k_{S_{j}}}{2}\left\{ \cos\left(\Delta\theta_{S}-\Delta\phi_{S_{j}}\right)-\cos\left(\Delta\theta_{S}-\Delta\phi_{S_{j}}+\Delta k_{S}z\right)\right\} \right],
\protect\end{array}\label{eq:ZVS}
\end{equation}
}}

where $\mathcal{C}_{p_{j}}^{S}=4g\Lambda_{j}\left|\alpha_{p_{j}}\right|\left|\alpha_{L_{1}}\right|\left|\alpha_{L_{2}}\right|\left|\gamma\right|$
is the real parameter controlling the depth of Zeno parameter and
phase difference parameter in Stokes-probe coupling is $\Delta\phi_{S_{j}}=(\phi_{S}-\phi_{p_{j}})$.
Also, phase mismatch parameter $\Delta k_{S_{j}}=(k_{S}-k_{p_{j}})$.
Note that $Z_{V}^{S}=-\stackrel[j=1]{2}{\sum}\mathcal{C}_{p_{j}}^{S}Z_{S_{j}}^{L}\left(\Delta k_{L_{j}}=-\Delta k_{S_{j}},\Delta\phi_{L_{j}}=-\Delta\phi_{S_{j}}\right)$.
As $\mathcal{C}_{p_{j}}^{S}$ only depends on $\gamma$, we can observe
quantum Zeno and anti-Zeno effects in partially stimulated cases when
$\beta=0=\delta\neq\gamma$. Further, we can observe that in phase
matching condition between Stokes and Stokes-probe interactions, we
obtain $\mathcal{Z}_{V}^{S}=-\frac{1}{4}\stackrel[j=1]{2}{\sum}\mathcal{C}_{p_{j}}^{S}z^{2}\cos\left(\Delta\theta_{S}-\Delta\phi_{S_{j}}\right)$.
Due to this symmetry it can be concluded that quantum Zeno (anti-Zeno)
effect in the phonon mode when the probe interacts with the Stokes
mode can be observed when the quantum anti-Zeno (Zeno) effect (after
the parameter substitution) was observed in the Stokes mode when the
probe was interacting with the pump mode.

\subsection{Case~III: Interaction of the probe with anti-Stokes mode only}

Interaction of the anti-Stokes mode with the probe can be obtained
as $\Omega_{l}\neq0$ and $\Gamma_{l}=0=\Lambda_{l}$. We study the
quantum Zeno and anti-Zeno effects in this case for Stoke and phonon
modes only as anti-Stokes mode is interacting with the probe. In this
case, Zeno parameter for the Stokes modes vanishes in the range of
the validity of the solution, i.e., $\begin{array}{lcl}
Z_{S}^{A} & = & 0\end{array}$. Here, $S$ in the subscript corresponds to Zeno parameter of Stokes
mode when the probe is interacting with the anti-Stokes mode, represented
by superscript $A$. It is worth mentioning here that the corresponding
parameter with short-length approximation technique is also fourth
order in rescaled length \cite{thun2002zeno-raman}. The Zeno parameter
for phonon mode in this case is 

\begin{equation}
\begin{array}{lcl}
Z_{V}^{A} & = & \stackrel[j=1]{2}{\sum}\mathcal{D}_{p_{j}}^{A}Z_{A_{j}}^{L}\left(\Delta k_{L_{j}}=-\Delta k_{A_{j}},\Delta\phi_{L_{j}}=-\Delta\phi_{A_{j}}\right),\end{array}\label{eq:ZVA}
\end{equation}
where $\mathcal{D}_{p_{j}}^{A}=\mathcal{C}_{p_{j}}^{S}\frac{\chi}{g}\frac{\Omega_{j}}{\Lambda_{j}}$
is the scaling factor, and the anti-Stokes-probe phase difference
parameter is $\Delta\phi_{A_{j}}=(\phi_{A}-\phi_{p_{j}})$. Also,
the phase mismatch parameter $\Delta k_{A_{j}}=(k_{A}-k_{p_{j}})$.
Here, we can observe quantum Zeno and anti-Zeno effects in partially
stimulated cases when $\beta=0=\delta\neq\gamma$. Further, we can
observe that in phase matching condition between anti-Stokes and anti-Stokes-probe
interactions, we obtain $\mathcal{Z}_{V}^{A}=-\frac{1}{4}\stackrel[j=1]{2}{\sum}\mathcal{D}_{p_{j}}^{A}z^{2}\cos\left(\Delta\theta_{A}+\Delta\phi_{A_{j}}\right)$.
This symmetry allows us to conclude that quantum Zeno (anti-Zeno)
effect in the phonon mode when the probe interacts with the anti-Stokes
mode can be observed (after the parameter substitution) when the quantum
Zeno (anti-Zeno) effect was observed in the anti-Stokes mode when
the probe was interacting with the pump mode.

\subsection{Case~IV: Interaction of a probe with Stokes and another probe with
anti-Stokes mode only}

Interaction of the Stokes mode with the first probe can be obtained
as $\Lambda_{1}\neq0$ and interaction of the anti-Stokes mode with
the second probe can be obtained as $\Omega_{2}\neq0$, whereas the
rest of the parameters are $\Gamma_{l}=0=\Omega_{1}=\Lambda_{2}$.
We study the quantum Zeno and anti-Zeno effects in this case for phonon
modes only as Stokes and anti-Stokes modes are interacting with the
probe. The Zeno parameter for phonon mode in this case is 
\begin{equation}
Z_{V}^{S,A}=-\mathcal{C}_{p_{1}}^{S}Z_{S_{1}}^{L}\left(\Delta k_{L_{1}}=-\Delta k_{S_{1}},\Delta\phi_{L_{1}}=-\Delta\phi_{S_{1}}\right)+\mathcal{D}_{p_{2}}^{A}Z_{A_{2}}^{L}\left(\Delta k_{L_{j}}=-\Delta k_{A_{j}},\Delta\phi_{L_{j}}=-\Delta\phi_{A_{j}}\right).\label{eq:ZVSA}
\end{equation}
Here, $V$ in the subscript corresponds to Zeno parameter of phonon
mode when the probe is interacting with the Stokes and anti-Stokes
modes, represented by superscript $S,A$. 

We can observe the quantum Zeno and anti-Zeno effects in partially
stimulated cases when $\beta=0=\delta\neq\gamma$. Further, we can
observe that in phase matching condition between Stokes and Stokes-probe
interactions, we obtain $\mathcal{Z}_{V}^{S,A}=-\frac{1}{4}\mathcal{C}_{p_{1}}^{S}z^{2}\cos\left(\Delta\theta_{S}-\Delta\phi_{S_{1}}\right)-\frac{1}{4}\mathcal{D}_{p_{j}}^{A}z^{2}\cos\left(\Delta\theta_{A}+\Delta\phi_{A_{2}}\right)$.
Due to this symmetry it can be concluded that both quantum Zeno and
anti-Zeno effects in the phonon mode can be observed when the probes
interact with the Stokes and anti-Stokes modes. 

\section{Photon Antibunching\label{sec:Photon-Antibunching}}

The information of single photon emission can be understood using
its photon statistics and is therefore of our central interest. The
single photon emission performance can be expressed in terms of its
second order intensity correlation function $g_{i}^{(2)}$. In order
to study bunching and anti-bunching properties of the radiation field
and phonon, we calculate the second-order correlation function \cite{perina-book}
\begin{equation}
g_{i}^{\left(2\right)}(z)=\frac{\left\langle i^{\dagger}(z)i^{\dagger}(z)i(z)i(z)\right\rangle }{\left\langle i^{\text{\dag}}(z)i(z)\right\rangle \left\langle i^{\text{\dag}}(z)i(z)\right\rangle },\label{secondordercorrelation1}
\end{equation}
where $i\in\left\{ a_{S},a_{V},a_{A}\right\} $ and $i\,(i^{\dagger})$
is the usual annihilation (creation) operator of the $i^{th}$ mode
of the radiation field. If $g_{i}^{\left(2\right)}>1$, the respective
light is thermal/chaotic in nature, and it shows the bunching phenomena.
Interestingly, the radiation field prepared in coherent state gives
rise to $g_{i}^{\left(2\right)}=1$, and hence the photons are unbunched
(i.e., neither showing bunching nor antibunching). Photon exhibits
antibunching for $g_{i}^{\left(2\right)}<1$. Equation (\ref{secondordercorrelation1})
can be rewritten as 

\begin{equation}
\begin{array}{lcl}
g_{i}^{(2)}-1 & = & \frac{\left\langle \Delta N_{i}\right\rangle ^{2}-\left\langle N_{i}\right\rangle }{\left\langle N_{i}\right\rangle ^{2}}\end{array}=\frac{D_{i}}{\left\langle N_{i}\right\rangle ^{2}},\label{secondordercorrelation2}
\end{equation}
where$\begin{array}{c}
D_{i}=\left\langle \Delta N_{i}\right\rangle ^{2}-\left\langle N_{i}\right\rangle \end{array}$is the difference between $\left\langle \Delta N_{i}\right\rangle ^{2}$
and $\left\langle N_{i}\right\rangle $ and the value of the numerator
$D_{i}$ determines (as the denominator is always positive) photon
bunching and antibunching. Precisely, $D_{i}>0,$$D_{i}=0$ and $D_{i}<0$
correspond photon bunching, unbunched and antibunching, respectively.
Using the above condition for various modes, we have
\begin{equation}
\begin{array}{lcl}
D_{S} & = & \frac{4g^{2}}{\Delta k_{S}^{2}}\left(1-\cos\Delta k_{s}z\right)\left|\alpha_{L_{1}}\right|^{2}\left|\alpha_{L_{2}}\right|^{2}\left|\beta\right|^{2}\end{array},\label{eq:antibunching-stokes}
\end{equation}

\begin{equation}
\begin{array}{lcl}
D_{V} & {\normalcolor {\normalcolor {\normalcolor =}}} & \frac{8g^{2}}{\Delta k_{S}^{2}}\left|\alpha_{L_{1}}\right|^{2}\left|\alpha_{L_{2}}\right|^{2}\left|\gamma\right|^{2}\left(1-\cos\Delta k_{s}z\right)+8\frac{\chi^{2}}{\Delta k_{A}^{2}}\left|\gamma\right|^{2}\left|\delta\right|^{2}\left(\left|\alpha_{L_{1}}\right|^{2}+\left|\alpha_{L_{2}}\right|^{2}+1\right)\\
{\normalcolor {\normalcolor {\normalcolor {\color{red}{\color{red}{\color{orange}}}}}}} & \times & \left(1-\cos\Delta k_{A}z\right){\normalcolor {\normalcolor +\frac{4g\chi}{\Delta k_{A}\Delta k_{S}}}{\normalcolor \left(\left|\alpha_{L_{1}}\right|^{2}+\left|\alpha_{L_{2}}\right|^{2}+1\right)\left|\beta\right|\left|\gamma\right|^{2}\left|\delta\right|}}\\
{\normalcolor {\normalcolor {\normalcolor {\color{red}}}}} & \times & \cos\left(\Delta\theta_{S}+\Delta\theta_{A}+\frac{\Delta k_{A}z+\Delta k_{S}z}{2}\right)\left[\cos\left(\frac{\Delta k_{S}z+\Delta k_{A}z}{2}\right)-\cos\left(\frac{\Delta k_{S}z-\Delta k_{A}z}{2}\right)\right]\\
{\normalcolor {\normalcolor {\normalcolor {\color{red}{\color{red}{\color{orange}}}}}}} & {\normalcolor {\normalcolor +}} & \frac{4g\chi}{\Delta k_{A}\Delta k_{S}\left(\Delta k_{S}+\Delta k_{A}\right)}\left|\beta\right|\left|\gamma\right|^{2}\left|\delta\right|\left(\left|\alpha_{L_{1}}\right|^{2}+\left|\alpha_{L_{2}}\right|^{2}+1\right)\left[\Delta k_{S}\cos\left(\frac{\Delta k_{S}z}{2}\right)\right.\\
{\normalcolor {\normalcolor {\normalcolor {\color{red}}}}} & \times & \left\{ \cos\left(\Delta\theta_{S}+\Delta\theta_{A}-\Delta k_{A}z-\frac{\Delta k_{S}z}{2}\right)-\cos\left(\Delta\theta_{S}+\Delta\theta_{A}-\frac{\Delta k_{S}z}{2}\right)\right\} \\
{\normalcolor {\normalcolor {\normalcolor {\color{red}{\color{red}{\color{orange}}}}}}} & {\normalcolor {\normalcolor +}} & \Delta k_{A}\cos\left(\frac{\Delta k_{A}z}{2}\right)\left\{ \cos\left(\Delta\theta_{S}+\Delta\theta_{A}-\frac{\Delta k_{A}z}{2}\right)-\cos\left(\Delta\theta_{S}+\Delta\theta_{A}-\Delta k_{S}z-\frac{\Delta k_{A}z}{2}\right)\right\} ,
\end{array}\label{eq:antibunching-phonon}
\end{equation}
 and
\begin{equation}
\begin{array}{lcl}
D_{A} & = & 0.\end{array}\label{eq:antibunching-anti-stokes}
\end{equation}

We can observe that for Stokes and anti-Stokes modes $D_{j}\geq0$.
Therefore, antibunching of photons in these two modes could not be
observed. Similarly, we failed to observe antibunching in the phonon
mode in the domain of the validity of the solution. Also, note that
in all these cases, there is no term dependent on $\Omega_{l}$, $\Gamma_{l}$,
and $\Lambda_{l}$. Therefore, in these cases, the couplings leading
to QZE and QAZE have no effect or equivalently the probes have no
effect on the single mode antibunching. 

In order to get the information about the bunching and antibunching
in the compound modes, we define the second order correlation function
for compound modes as
\begin{equation}
g_{ij}^{\left(2\right)}(z)=\frac{\left\langle i^{\text{\dag}}(z)j^{\text{\dag}}(z)j(z)i(z)\right\rangle }{\left\langle i^{\text{\dag}}(z)i(z)\right\rangle \left\langle j^{\text{\dag}}(z)j(z)\right\rangle }\label{eq:secondordercorrelation-twomode}
\end{equation}
where $i$ and $j$ are the usual annihilation (creation) operator
with $i,j\in\left\{ a_{S},a_{V},a_{A}\right\} $ and $i\neq j.$ Equation
(\ref{eq:secondordercorrelation-twomode}) can be rewritten in the
following form:
\begin{equation}
\begin{array}{lcl}
g_{ij}^{\left(2\right)} & = & 1+\frac{\left(\Delta N_{ij}\right)^{2}}{N_{i}N_{j}}=1+\frac{D_{ij}}{N_{i}N_{j}}\end{array},\label{eq:expression-antibunching}
\end{equation}
where $D_{ij}(z)=$ $\left(\Delta N_{ij}\right)^{2}=\left\langle i^{\text{\dag}}(z)j^{\text{\dag}}(z)j(z)i(z)\right\rangle -\left\langle i^{\text{\dag}}(z)i(z)\right\rangle \left\langle j^{\text{\dag}}(z)j(z)\right\rangle $
and again it determines the bunching and antibunching properties of
the compound mode. In Stokes-phonon mode, we have

\begin{equation}
\begin{array}{lcl}
D_{SV} & {\normalcolor =} & 2\frac{g^{2}}{\left(\Delta k_{S}\right)^{2}}\left[\left|\alpha_{L_{1}}\right|\left|\alpha_{L_{2}}\right|^{2}\left(2\left|\gamma\right|^{2}+1\right)-\left(\left|\alpha_{L_{1}}\right|^{2}+1\right)\left|\beta\right|^{2}\left|\gamma\right|^{2}+\left(\left|\alpha_{L_{1}}\right|^{2}-\left|\gamma\right|^{2}\right)\left|\alpha_{L_{2}}\right|^{2}\left|\beta\right|^{2}\right]\\
 & \times & \left(1-\cos\Delta k_{S}z\right)+\frac{2g}{\Delta k_{S}}\left|\alpha_{L_{1}}\right|\left|\alpha_{L_{2}}\right|\left|\beta\right|\left|\gamma\right|\left[\cos\Delta\theta_{S}-\cos\left(\Delta\theta_{S}+\Delta k_{S}z\right)\right]+\frac{4g\chi\left|\beta\right|\left|\gamma\right|^{2}\left|\delta\right|}{\left(\Delta k_{S}+\Delta k_{A}\right)\Delta k_{S}\Delta k_{A}}\\
 & \times & \left(\left|\alpha_{L_{1}}\right|^{2}+\left|\alpha_{L_{2}}\right|^{2}+1\right)\left[\Delta k_{S}\left\{ \cos\left(\Delta\theta_{S}+\Delta\theta_{A}-\Delta k_{S}z-\Delta k_{A}z\right)-\cos\left(\Delta\theta_{S}+\Delta\theta_{A}-\Delta k_{S}z\right)\right\} \right.\\
 & - & \left.\Delta k_{A}\left\{ \cos\left(\Delta\theta_{S}+\Delta\theta_{A}-\Delta k_{S}z\right)-\cos\left(\Delta\theta_{S}+\Delta\theta_{A}\right)\right\} \right]+\frac{4g\varLambda_{1}\left|\alpha_{P_{1}}\right|\left|\alpha_{L_{1}}\right|\left|\alpha_{L_{2}}\right|\left|\gamma\right|}{\Delta k_{S}\Delta k_{S_{1}}}\\
 & \times & \cos\left\{ \Delta\theta_{S}-\Delta\phi_{s_{1}}-\frac{\left(\Delta k_{S_{1}}-\Delta k_{S}\right)z}{2}\right\} \left[\cos\frac{\left(\Delta k_{S_{1}}-\Delta k_{S}\right)z}{2}-\cos\frac{\left(\Delta k_{S_{1}}+\Delta k_{S}\right)z}{2}\right]+\frac{4g\varLambda_{2}\left|\alpha_{P_{2}}\right|\left|\alpha_{L_{1}}\right|\left|\alpha_{L_{2}}\right|\left|\gamma\right|}{\Delta k_{S}\Delta k_{S_{2}}}\\
 & \times & \cos\left\{ \Delta\theta_{S}-\Delta\phi_{s_{2}}-\frac{\left(\Delta k_{S_{2}}-\Delta k_{S}\right)z}{2}\right\} \left[\cos\frac{\left(\Delta k_{S_{2}}-\Delta k_{S}\right)z}{2}-\cos\frac{\left(\Delta k_{S_{2}}+\Delta k_{S}\right)z}{2}\right]+\frac{g\varGamma_{1}\left|\alpha_{P_{1}}\right|\left|\alpha_{L_{2}}\right|\left|\beta\right|\left|\gamma\right|}{\Delta k_{S}\left(\Delta k_{S}+\Delta k_{L_{1}}\right)\Delta k_{L_{1}}}\\
 & \times & \left[\Delta k_{S}\left\{ \cos\left(\Delta\theta_{S}+\Delta\phi_{L_{j}}-\Delta k_{S}z-\Delta k_{L_{1}}z\right)-\cos\left(\Delta\theta_{S}+\Delta\phi_{L_{j}}-\Delta k_{S}z\right)\right\} \right.\\
 & - & \left.\Delta k_{L_{1}}\left\{ \cos\left(\Delta\theta_{S}+\Delta\phi_{L_{j}}-\Delta k_{S}z\right)-\cos\left(\Delta\theta_{S}+\Delta\phi_{L_{j}}\right)\right\} \right]+\frac{g\varGamma_{2}\left|\alpha_{P_{2}}\right|\left|\alpha_{L_{1}}\right|\left|\beta\right|\left|\gamma\right|}{\Delta k_{S}\left(\Delta k_{S}+\Delta k_{L_{2}}\right)\Delta k_{L_{2}}}\\
 & \times & \left[\Delta k_{S}\left\{ \cos\left(\Delta\theta_{S}+\Delta\phi_{L_{j}}-\Delta k_{S}z-\Delta k_{L_{2}}z\right)-\cos\left(\Delta\theta_{S}+\Delta\phi_{L_{j}}-\Delta k_{S}z\right)\right\} \right.\\
 & - & \left.\Delta k_{L_{2}}\left\{ \cos\left(\Delta\theta_{S}+\Delta\phi_{L_{j}}-\Delta k_{S}z\right)-\cos\left(\Delta\theta_{S}+\Delta\phi_{L_{j}}\right)\right\} \right]\\
 & + & \frac{4g\chi\left|\alpha_{L_{1}}\right|^{2}\left|\alpha_{L_{2}}\right|^{2}\left|\beta\right|\left|\delta\right|}{\Delta k_{S}\Delta k_{A}}\cos\left(\Delta\theta_{S}-\Delta\theta_{A}+\frac{\left(\Delta k_{S}-\Delta k_{A}\right)z}{2}\right)\left[\cos\frac{\left(\Delta k_{S}-\Delta k_{A}\right)z}{2}-\cos\frac{\left(\Delta k_{S}+\Delta k_{A}\right)z}{2}\right],
\end{array}\label{eq:antibunching-stokes-phonon}
\end{equation}

\begin{figure}
\centering{}\includegraphics[scale=0.6]{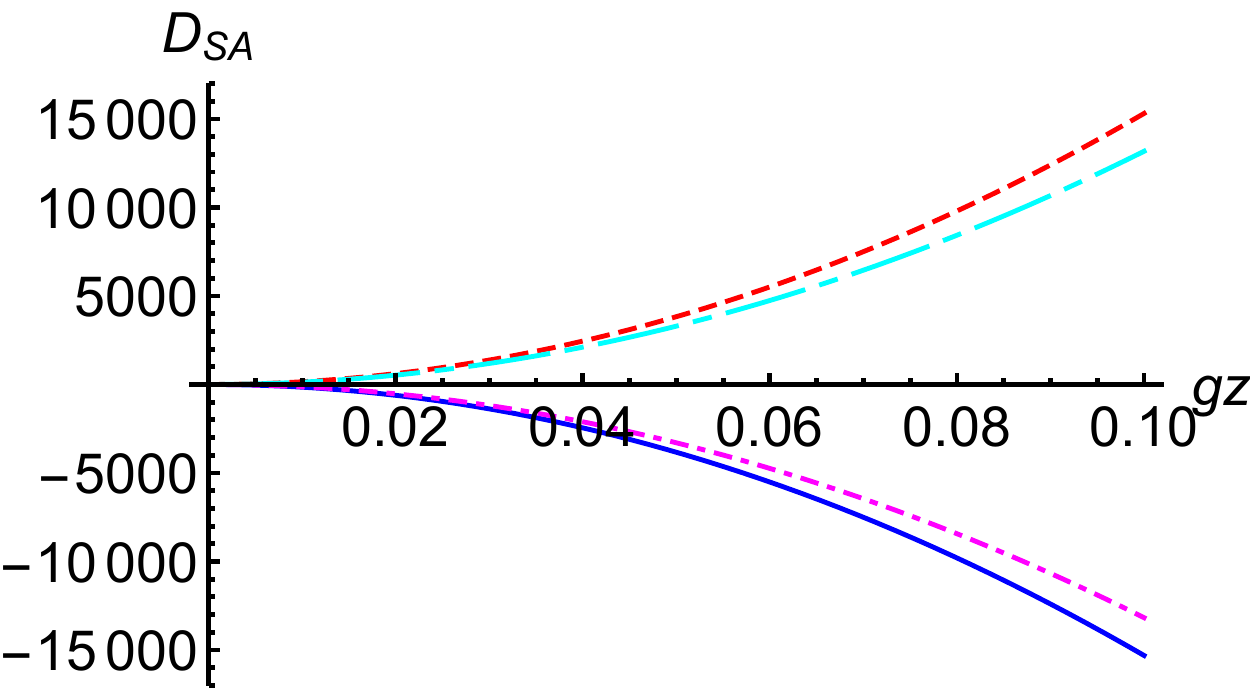}\caption{\label{fig:DSA}(Color online) Spatial evolution of antibunching in
compound Stokes--anti-Stokes mode $D_{SA}$ for $\Delta\theta_{S}=\Delta\theta_{A}=0$
and phase mismatches $\Delta k_{S}=10^{-2}g$ and $\Delta k_{A}=1.1\Delta k_{S}$
(in the smooth (blue) line). This antibunching disappears for $\Delta\theta_{S}-\Delta\theta_{A}=\pi$
(in the dashed (red) line). On the other hand, antibunching could
be observed for $\Delta k_{A}=0.9\Delta k_{S}$ if $\Delta\theta_{S}-\Delta\theta_{A}=\pi$
(in the dot-dashed (magenta) line) but not when $\Delta\theta_{S}=\Delta\theta_{A}=0$
(in the dot-dashed (cyan) line). }
\end{figure}

Interestingly, in contrast to single mode parameters $D_{i},$ here,
$D_{SV}$ depends on the coupling parameters. Thus, probing affects
$D_{SV}.$ Though probing affects $D_{SV},$ we failed to observe
antibunching for Stokes-phonon mode in domain of the validity of the
perturbative solution. To investigate the possibility of observing
intermodal antibunching involving Stokes and anti-stokes mode, we
compute $D_{SA}$ as
\begin{equation}
\begin{array}{lcl}
D_{SA} & = & -\frac{2g\chi\left|\alpha_{L_{1}}\right|^{2}\left|\alpha_{L_{2}}\right|^{2}\left|\beta\right|\left|\delta\right|}{\Delta k_{S}\Delta k_{A}}\left[2\cos\left(\Delta\theta_{S}-\Delta\theta_{A}+\frac{\left(\Delta k_{S}-\Delta k_{A}\right)z}{2}\right)\left[\cos\frac{\left(\Delta k_{S}-\Delta k_{A}\right)z}{2}-\cos\frac{\left(\Delta k_{S}+\Delta k_{A}\right)z}{2}\right]\right.\\
 & + & \frac{1}{\left(\Delta k_{S}-\Delta k_{A}\right)}\left\{ \Delta k_{S}\cos\left(\Delta\theta_{S}-\Delta\theta_{A}+\Delta k_{S}z+\Delta k_{A}z\right)-\Delta k_{A}\cos\left(\Delta\theta_{S}-\Delta\theta_{A}\right)\right\} -\cos\left(\Delta\theta_{S}-\Delta\theta_{A}+\Delta k_{S}z\right)
\end{array}\label{eq:antibuncing-stokes-anti-stokes}
\end{equation}
and found that the antibunching parameter does not depend on the parameters
of the probes. Clearly, the compound Stokes--anti-Stokes antibunching
depends only on $\Delta\theta_{S}-\Delta\theta_{A}$ and phase mismatches
$\Delta k_{S}$ and $\Delta k_{A}$. In Fig. \ref{fig:DSA}, we observed
antibunching in Stokes-anti-Stokes mode for $\Delta k_{S}>\Delta k_{A}$,
which disappears with change in the relative phase difference $\Delta\theta_{S}-\Delta\theta_{A}$.
However, an opposite effect is observed due to change in the relative
phase difference $\Delta\theta_{S}-\Delta\theta_{A}$ for the choice
of $\Delta k_{S}<\Delta k_{A}$.

Finally, from the following expression
\begin{equation}
\begin{array}{lcl}
D_{VA} & = & -2\frac{\chi^{2}}{\left(\Delta k_{A}\right)^{2}}\left(\left|\alpha_{L_{1}}\right|^{2}+\left|\alpha_{L_{2}}\right|^{2}+1\right)\left|\gamma\right|^{2}\left|\delta\right|^{2}\left(1-\cos\Delta k_{A}z\right)\end{array}\label{eq:antibunching-phonon-anti-stokes}
\end{equation}
we can observe that for phonon-anti-Stokes mode $D_{VA}\leq0$ and
thus intermodal antibunching of bosons is present. Note that antibunching
in the pump modes and coupled pump-pump mode in the hyper-Raman processes
is already illustrated \cite{hyper-R,Gerry}. Here, we have not discussed
the correponding cases. Hence it is clear from equations (\ref{eq:antibunching-stokes})-(\ref{eq:antibunching-phonon-anti-stokes}),
the expression of antibunching effect for single modes and the coupled
modes of system are free from the interaction of probe and hence,
we may conclude that there is no evidence of any direct relation of
photon antibunching with QZE (QAZE).

\section{Conclusions\label{sec:Conclusions}}

We have investigated the possibility of observing QZE and QAZE in
a nonlinear waveguide undergoing non-degenerate hyper-Raman processes
coupled to two linear waveguides. Specifically, in this work, in the
context of QZE and QAZE, we have studied the impact of the presence
of two probe waveguides on the dynamics of a hyper-Raman active waveguide.
While doing so, the enhancement (diminution) of the evolution of the
hyper-Raman processes due to interaction with the probe waveguides via
evanescent waves is viewed as QZE (QAZE). Further, we have tried to
understand the relationship between antibunching (as a representative
nonclassical phenomenon) and QZE/QAZE, but the study has not revealed
any specific relation. Interestingly, QZE and QAZE were reported earlier
in various non-optical \cite{QZ-no,QZ-no2,QZ-no3} and optical \cite{Rechacek-2001-zrno-coupler,thun2002zeno-raman,chi2-chi1-spie,chi2-chi2,rehacek2000zeno-coupler,down-conversion-perina,down-conversion-perina2,parametric-down-conversion-anti-zeno,perina-zeno-parametric-dc,All-optical-zeno-agarwal}
systems, but in most of those studies on optical systems, the pump
mode(s) was considered to be strong to circumvent the complexity of
a fully quantum approach. In contrast to those works, here we have
adopted a completely quantum treatment involving Sen-Mandal technique
as was done in \cite{moumita-zeno-1}. Specifically, a perturbative
operator solution is obtained using Sen-Mandal technique, and the
same is used to investigate the possibility of observing QZE, QAZE
and antibunching in different scenarios. The investigation has lead
to various interesting observations as summarized below.

When the interaction of probe modes is considered with the pump mode
only, we have observed QZE and QAZE in both phase match and phase
mismatch conditions for Stokes, anti-Stokes and phonon modes for the
stimulated cases, but QZE and QAZE are not observed for spontaneous
case. Further, the Zeno parameter is observed to be consistent with
the constant of motion. In a different scenario, when we consider
that the probe modes interact with the Stokes mode, then the Zeno
parameter is observed to disappear for anti-Stokes mode and to yield
a non-zero value for phonon mode only. Similarly, when the interaction
of the probe modes is considered with anti-Stokes mode, the Zeno parameter
vanishes for Stokes mode. However, in this situation, we have observed
the signature of Zeno effect in phonon mode for partially stimulated
process. Thus, it is concluded that QZE is observed in phonon mode
when probes interact with Stokes/anti-Stokes mode and QZE (QAZE) is
observed in Stokes/anti-Stokes mode when probes interact with the
pump mode. Finally, it was considered that one probe is interacting
with Stokes mode and the other one is interacting with the anti-Stokes
mode. In this scenario, for both stimulated as well as partially stimulated
cases, we have observed the presence of QZE (QAZE) in phonon mode
only. Though the central focus of the present paper remained to discuss
feasibility of observing QZE and QAZE in a hyper-Raman active waveguide,
we also attempted to investigate the role of probing on the quantum
photon statistics and to check whether there exists any relation between
antibunching and QZE/QAZE which originates due to coupling (interaction).
Interestingly, we found that the antibunching parameters for single
mode and inter-modal cases are usually independent of the coupling
parameters, i.e., the coupling constants quantifying interaction with
the probe. Specifically, single-mode photon antibunching phenomenon
is not observed in Stokes, anti-Stokes mode and phonon mode. Further,
the antibunching parameters of all these three modes are found to
be independent of $\Omega_{l},$$\Gamma_{l},$and $\Lambda_{l}$.
Hence, no effect of probing is observed on antibunching in these modes,
hinting that the QZE and antibuncing are independent processes. Thus
it may be concluded as there is no correlation between photon antibunching
and QZE (QAZE). Subsequently, antibunching in coupled mode is studied
in detail, and for the Stokes-anti-Stokes couple mode, antibunching
is observed for a proper choice of phase differences and phase mismatch.
Similar investigation revealed that intermodal antibunching parameter
for Phonon-anti-Stokes mode always show antibunching (does not depend
on the choice of phase difference and phase mismatch). However, as
observed in phonon mode, corresponding antibunching parameters (i.e.,
$D_{SA}$ and $D_{AV}$) are found to be independent of coupling constants
associated with the probe, and thus observed antibunching is found
to be independent of QZE. Finally, it is found that intermodal antibunching
parameter $D_{SV}$ involving Stokes and phonon modes are function
of coupling parameters, but it does not lead to antibuncing at any
point where the solution obtained here can be considered as valid
solution. 

Before we conclude this work, we would like to note that collapse
on measurement postulate distinguishes quantum mechanics from classical
theories allowing superposition, and many of the advantages of quantum
world (like, in quantum cryptography measurement performed for eavesdropping
leaves detectable trace) and troubles of quantum world (like the collapse/modification
of a quantum state prepared for quantum computing due to unwanted
interaction with the environment) are due to this nonclassical (collapse
on measurement) feature of quantum world. In fact, QZE and QAZE are
also manifestation of this feature and these can be observed even
in other nonclassical theories containing this postulate. Thus, it
is very general, in nature and naturally following the methodology
used here, QZE and QAZE can be observed in various physical systems
other than an optical system performing hyper-Raman processes studied
here. Further, as the system of our interest is easily realizable
using photonic crystals and optical fibers \cite{Dcoup1,Dcoup2,kishore2014co-coupler},
we hope that the results reported here will be realized soon and would
find applications in achieving desired quantum control useful in realizing
some exciting ideas of quantum computation, communication and metrology. 

\section*{Acknowledgment}

AP acknowledges the support from the QUEST scheme of the Interdisciplinary
Cyber-Physical Systems (ICPS) program of the Department of Science
and Technology (DST), India, Grant No.: DST/ICPS/QuST/Theme-1/2019/14
(Q80).

\section*{Appendix: A\label{sec:Appendix:-A}}

The assumed solution of Eq. (\ref{equationofmotion}) using Sen-Mandal
approach is obtained up to the quadratic term of the interaction constants
($g,$ $\chi,$ $\Gamma_{1}$, and $\Gamma_{2}$) under weak pump
approximation. These assumed solutions, i.e., the spatial evolution
of all field and phonon modes are in the following form: 
\begin{equation}
\begin{array}{lcl}
a_{p_{1}}(z) & = & f_{1}a_{p_{1}}(0)+f_{2}a_{L_{1}}(0)+f_{3}a_{S}(0)+f_{4}a_{A}(0)+f_{5}a_{L_{2}}^{\dagger}(0)a_{S}(0)a_{V}(0)+f_{6}a_{L_{2}}^{\dagger}(0)a_{V}^{\dagger}(0)a_{A}(0)\\
 & + & f_{7}a_{L_{1}}(0)a_{L_{2}}(0)a_{V}^{\dagger}(0)+f_{8}a_{p_{2}}(0)+f_{9}a_{L_{1}}(0)a_{L_{2}}(0)a_{V}(0)+f_{10}a_{p_{2}}(0)+f_{11}a_{p_{1}}(0)\\
 & + & f_{12}a_{p_{1}}(0)+f_{13}a_{p_{1}}(0),\\
a_{p_{2}}(z) & = & g_{1}a_{p_{2}}(0)+g_{2}a_{L_{2}}(0)+g_{3}a_{S}(0)+g_{4}a_{A}(0)+g_{5}a_{L_{1}}^{\dagger}(0)a_{S}(0)a_{V}(0)+g_{6}a_{L_{1}}^{\dagger}(0)a_{V}^{\dagger}(0)a_{A}(0)\\
 & + & g_{7}a_{L_{1}}(0)a_{L_{2}}(0)a_{V}^{\dagger}(0)+g_{8}a_{p_{1}}(0)+g_{9}a_{L_{1}}(0)a_{L_{2}}(0)a_{V}(0)+g_{10}a_{p_{1}}(0)+g_{11}a_{p_{2}}(0)\\
 & + & g_{12}a_{p_{2}}(0)+g_{13}a_{p_{2}}(0),\\
a_{L_{1}}(z) & = & h_{1}a_{L_{1}}(0)+h_{2}a_{L_{2}}^{\dagger}(0)a_{S}(0)a_{V}(0)+h_{3}a_{L_{2}}^{\dagger}(0)a_{V}^{\dagger}(0)a_{A}(0)+h_{4}a_{p_{1}}(0)\\
 & + & h_{5}a_{L_{1}}(0)a_{S}(0)a_{V}^{2}(0)a_{A}^{\dagger}(0)+h_{6}a_{L_{1}}^{\dagger}(0)a_{L_{2}}^{\dagger2}(0)a_{S}(0)a_{A}(0)\\
 & + & h_{7}a_{L_{1}}(0)a_{S}^{\dagger}(0)a_{V}^{\dagger2}(0)a_{A}(0)+h_{8}a_{p_{1}}(0)a_{L_{2}}^{\dagger}(0)a_{V}(0)+h_{9}a_{p_{2}}(0)a_{L_{2}}^{\dagger}(0)a_{V}(0)\\
 & + & h_{10}a_{p_{2}}^{\dagger}(0)a_{S}(0)a_{V}(0)+h_{11}a_{p_{1}}(0)a_{L_{2}}^{\dagger}(0)a_{V}^{\dagger}(0)+h_{12}a_{p_{2}}(0)a_{L_{2}}^{\dagger}(0)a_{V}^{\dagger}(0)\\
 & + & h_{13}a_{p_{2}}^{\dagger}(0)a_{V}^{\dagger}(0)a_{A}(0)+h_{14}a_{S}(0)+h_{15}a_{A}(0)+h_{16}a_{L_{1}}(0)a_{S}^{\dagger}(0)a_{S}(0)a_{V}^{\dagger}(0)a_{V}(0)\\
 & + & h_{17}a_{L_{1}}(0)a_{L_{2}}^{\dagger}(0)a_{L_{2}}(0)a_{S}^{\dagger}(0)a_{S}(0)+h_{18}a_{L_{1}}(0)a_{L_{2}}^{\dagger}(0)a_{L_{2}}(0)a_{V}(0)a_{V}^{\dagger}(0)\\
 & + & h_{19}a_{L_{1}}(0)a_{V}^{\dagger}(0)a_{V}(0)a_{A}^{\dagger}(0)a_{A}(0)+h_{20}a_{L_{1}}(0)a_{L_{2}}^{\dagger}(0)a_{L_{2}}(0)a_{V}^{\dagger}(0)a_{V}(0)\\
 & + & h_{21}a_{L_{1}}(0)a_{L_{2}}(0)a_{L_{2}}^{\dagger}(0)a_{A}^{\dagger}(0)a_{A}(0)+h_{22}a_{L_{1}}(0),\\
a_{L_{2}}(z) & = & J_{1}a_{L_{2}}(0)+J_{2}a_{L_{1}}^{\dagger}(0)a_{S}(0)a_{V}(0)+J_{3}a_{L_{1}}^{\dagger}(0)a_{V}^{\dagger}(0)a_{A}(0)+J_{4}a_{p_{2}}(0)\\
 & + & J_{5}a_{L_{2}}(0)a_{S}(0)a_{V}^{2}(0)a_{A}^{\dagger}(0)+J_{6}a_{L_{1}}^{\dagger2}(0)a_{L_{2}}^{\dagger}(0)a_{S}(0)a_{A}(0)+J_{7}a_{L_{2}}(0)a_{S}^{\dagger}(0)a_{V}^{\dagger2}(0)a_{A}(0)\\
 & + & J_{8}a_{p_{1}}(0)a_{L_{1}}^{\dagger}(0)a_{V}(0)+J_{9}a_{p_{2}}(0)a_{L_{1}}^{\dagger}(0)a_{V}(0)+J_{10}a_{p_{1}}^{\dagger}(0)a_{S}(0)a_{V}(0)\\
 & + & J_{11}a_{p_{1}}(0)a_{L_{1}}^{\dagger}(0)a_{V}^{\dagger}(0)+J_{12}a_{p_{2}}a_{L_{1}}^{\dagger}(0)a_{V}^{\dagger}(0)+J_{13}a_{p_{1}}^{\dagger}(0)a_{V}^{\dagger}(0)a_{A}(0)\\
 & + & J_{14}a_{S}(0)+J_{15}a_{A}(0)+J_{16}a_{L_{2}}(0)a_{S}^{\dagger}(0)a_{S}(0)a_{V}^{\dagger}(0)a_{V}(0)\\
 & + & J_{17}a_{L_{1}}^{\dagger}(0)a_{L_{1}}(0)a_{L_{2}}(0)a_{S}^{\dagger}(0)a_{S}(0)+J_{18}a_{L_{1}}^{\dagger}(0)a_{L_{1}}(0)a_{L_{2}}(0)a_{V}(0)a_{V}^{\dagger}(0)\\
 & + & J_{19}a_{L_{2}}(0)a_{V}^{\dagger}(0)a_{V}(0)a_{A}^{\dagger}(0)a_{A}(0)+J_{20}a_{L_{1}}^{\dagger}(0)a_{L_{1}}(0)a_{L_{2}}(0)a_{V}^{\dagger}(0)a_{V}(0)\\
 & + & J_{21}a_{L_{1}}(0)a_{L_{1}}^{\dagger}(0)a_{L_{2}}(0)a_{A}^{\dagger}(0)a_{A}(0)+J_{22}a_{L_{2}}(0),\\
a_{S}(z) & = & l_{1}a_{S}(0)+l_{2}a_{L_{1}}(0)a_{L_{2}}(0)a_{V}^{\dagger}(0)+l_{3}a_{p_{1}}(0)+l_{4}a_{p_{2}}(0)+l_{5}a_{L_{1}}^{2}(0)a_{L_{2}}^{2}(0)a_{A}^{\dagger}(0)\\
 & + & l_{6}a_{L_{1}}(0)a_{L_{1}}^{\dagger}(0)a_{V}^{\dagger2}(0)a_{A}(0)+l_{7}a_{L_{2}}^{\dagger}(0)a_{L_{2}}(0)a_{V}^{\dagger2}(0)a_{A}(0)+l_{8}a_{p_{1}}(0)a_{L_{2}}(0)a_{V}^{\dagger}(0)\\
 & + & l_{9}a_{p_{2}}(0)a_{L_{1}}(0)a_{V}^{\dagger}(0)+l_{10}a_{L_{1}}(0)+l_{11}a_{L_{2}}(0)+l_{12}a_{A}(0)+l_{13}a_{A}(0)\\
 & + & l_{14}a_{L_{1}}(0)a_{L_{1}}^{\dagger}(0)a_{L_{2}}(0)a_{L_{2}}^{\dagger}(0)a_{S}(0)+l_{15}a_{L_{1}}(0)a_{L_{1}}^{\dagger}(0)a_{S}(0)a_{V}(0)a_{V}^{\dagger}(0)\\
 & + & l_{16}a_{L_{2}}^{\dagger}(0)a_{L_{2}}(0)a_{S}(0)a_{V}(0)a_{V}^{\dagger}(0)+l_{17}a_{S}(0)+l_{18}a_{S}(0),\\
a_{V}(z) & = & m_{1}a_{V}(0)+m_{2}a_{L_{1}}(0)a_{L_{2}}(0)a_{S}^{\dagger}(0)+m_{3}a_{L_{1}}^{\dagger}(0)a_{L_{2}}^{\dagger}(0)a_{A}(0)+m_{4}a_{L_{1}}^{\dagger}(0)a_{L_{1}}(0)a_{S}^{\dagger}(0)a_{V}^{\dagger}(0)a_{A}(0)\\
 & + & m_{5}a_{L_{2}}^{\dagger}(0)a_{L_{2}}(0)a_{S}^{\dagger}(0)a_{V}^{\dagger}(0)a_{A}(0)+m_{6}a_{S}^{\dagger}(0)a_{V}^{\dagger}(0)a_{A}(0)+m_{7}a_{p_{1}}(0)a_{L_{2}}(0)a_{S}^{\dagger}(0)+m_{8}a_{p_{2}}(0)a_{L_{1}}(0)a_{S}^{\dagger}(0)\\
 & + & m_{9}a_{p_{1}}^{\dagger}(0)a_{L_{1}}(0)a_{L_{2}}(0)+m_{10}a_{p_{2}}^{\dagger}(0)a_{L_{1}}(0)a_{L_{2}}(0)+m_{11}a_{p_{1}}^{\dagger}(0)a_{L_{2}}^{\dagger}(0)a_{A}(0)+m_{12}a_{p_{2}}^{\dagger}(0)a_{L_{1}}^{\dagger}(0)a_{A}(0)\\
 & + & m_{13}a_{p_{1}}(0)a_{L_{1}}^{\dagger}(0)a_{L_{2}}^{\dagger}(0)+m_{14}a_{p_{2}}(0)a_{L_{1}}^{\dagger}(0)a_{L_{2}}^{\dagger}(0)+m_{15}a_{L_{1}}(0)a_{L_{1}}^{\dagger}(0)a_{L_{2}}(0)a_{L_{2}}^{\dagger}(0)a_{V}(0)\\
 & + & m_{16}a_{L_{1}}(0)a_{L_{1}}^{\dagger}(0)a_{S}(0)a_{S}^{\dagger}(0)a_{V}(0)+m_{17}a_{L_{2}}^{\dagger}(0)a_{L_{2}}(0)a_{S}(0)a_{S}^{\dagger}(0)a_{V}(0)\\
 & + & m_{18}a_{L_{1}}^{\dagger}(0)a_{L_{1}}(0)a_{L_{2}}^{\dagger}(0)a_{L_{2}}(0)a_{V}(0)+m_{19}a_{L_{1}}^{\dagger}(0)a_{L_{1}}(0)a_{V}(0)a_{A}^{\dagger}(0)a_{A}(0)\\
 & + & m_{20}a_{L_{2}}(0)a_{L_{2}}^{\dagger}(0)a_{V}(0)a_{A}^{\dagger}(0)a_{A}(0),\\
a_{A}(z) & = & n_{1}a_{A}(0)+n_{2}a_{L_{1}}(0)a_{L_{2}}(0)a_{V}(0)+n_{3}a_{p_{1}}(0)+n_{4}a_{p_{2}}(0)+n_{5}a_{L_{1}}^{2}(0)a_{L_{2}}^{2}(0)a_{S}^{\dagger}(0)\\
 & + & n_{6}a_{L_{1}}(0)a_{L_{1}}^{\dagger}(0)a_{S}(0)a_{V}^{2}(0)+n_{7}a_{L_{2}}^{\dagger}(0)a_{L_{2}}(0)a_{S}(0)a_{V}^{2}(0)+n_{8}a_{p_{1}}(0)a_{L_{2}}(0)a_{V}(0)\\
 & + & n_{9}a_{p_{2}}(0)a_{L_{1}}(0)a_{V}(0)+n_{10}a_{L_{1}}(0)+n_{11}a_{L_{2}}(0)+n_{12}a_{S}(0)+n_{13}a_{S}(0)+\\
 & + & n_{14}a_{L_{1}}(0)a_{L_{1}}^{\dagger}(0)a_{L_{2}}(0)a_{L_{2}}^{\dagger}(0)a_{A}(0)+n_{15}a_{L_{1}}(0)a_{L_{1}}^{\dagger}(0)a_{V}^{\dagger}(0)a_{V}(0)a_{A}(0)\\
 & + & n_{16}a_{L_{2}}^{\dagger}(0)a_{L_{2}}(0)a_{V}^{\dagger}(0)a_{V}(0)a_{A}(0)+n_{17}a_{A}(0)+n_{18}a_{A}(0).
\end{array}\label{eq:assumedsolution}
\end{equation}
From these equations, we can easily observe that at $z=0,$i.e., in
the absence of interaction, the parameters $x_{k\neq1}=0\,\forall k$
and $x_{1}=1,$where $x\in\{f,g,\,h,\,J,\,l,\,m,\,n\}.$ In contrast,
when interacton happens, i.e., when $z\neq0,$ then the parameters
$x_{k}\forall k$ would be functions of $z$ as the above trial solutions
(\ref{eq:assumedsolution}) are obtained by using 

\begin{equation}
\begin{array}{ccc}
a_{k^{\prime}}\left(z\right) & = & \exp\left(iGz\right)a_{k^{\prime}}\left(0\right)\exp\left(-iGz\right)\end{array}:\,k^{\prime}\in\{p_{1},p_{2},L_{1},L_{2},S,V,A\}.
\end{equation}
The functional form of all coefficients are 
\begin{equation}
\begin{array}{lcl}
l_{1} & = & e^{izk_{S}}\\
l_{2} & = & \frac{e^{izk_{S}}\left(1-e^{-iz\Delta k_{S}}\right)g}{\Delta k_{S}}\\
l_{3} & = & \frac{e^{izk_{S}}\left(1-e^{-iz\Delta k_{S_{1}}}\right)\varLambda_{1}}{\Delta k_{S_{1}}}\\
l_{4} & = & \frac{e^{izk_{S}}\left(1-e^{-iz\Delta k_{S_{2}}}\right)\varLambda_{2}}{\Delta k_{S_{2}}}\\
l_{5} & =- & \frac{e^{izk_{S}}\left(\left(\Delta k_{S}-\Delta k_{A}\right)e^{-iz\Delta k_{S}}-\Delta k_{S}e^{-iz\left(\Delta k_{S}-\Delta k_{A}\right)}+\Delta k_{A}\right)g\chi}{\Delta k_{S}\Delta k_{A}\left(\Delta k_{S}-\Delta k_{A}\right)}\\
l_{6} & = & l_{7}=\frac{e^{izk_{S}}\left(\Delta k_{S}e^{-iz\left(\Delta k_{S}+\Delta k_{A}\right)}-\left(\Delta k_{S}+\Delta k_{A}\right)e^{-iz\Delta k_{S}}+\Delta k_{A}\right)g\chi}{\left(\Delta k_{S}+\Delta k_{A}\right)\Delta k_{S}\Delta k_{A}}\\
l_{8} & = & \frac{e^{izk_{S}}\left(\Delta k_{S}e^{-iz\left(\Delta k_{S}+\Delta k_{L_{1}}\right)}-\left(\Delta k_{S}+\Delta k_{L_{1}}\right)e^{-iz\Delta k_{S}}+\Delta k_{L_{1}}\right)g\varGamma_{1}}{\Delta k_{S}\left(\Delta k_{S}+\Delta k_{L_{1}}\right)\Delta k_{L_{1}}}\\
l_{9} & = & \frac{e^{izk_{S}}\left(\Delta k_{S}e^{-iz\left(\Delta k_{S}+\Delta k_{L_{2}}\right)}-\left(\Delta k_{S}+\Delta k_{L_{2}}\right)e^{-iz\Delta k_{S}}+\Delta k_{L_{2}}\right)g\varGamma_{2}}{\Delta k_{S}\left(\Delta k_{S}+\Delta k_{L_{2}}\right)\Delta k_{L_{2}}}\\
l_{10} & = & \frac{e^{izk_{S}}\left(\left(\Delta k_{S_{1}}-\Delta k_{L_{1}}\right)e^{-iz\Delta k_{S_{1}}}-\Delta k_{S_{1}}e^{-iz\left(\Delta k_{S_{1}}-\Delta k_{L_{1}}\right)}+\Delta k_{L_{1}}\right)\varGamma_{1}\varLambda_{1}}{\left(\Delta k_{S_{1}}-\Delta k_{L_{1}}\right)\Delta k_{S_{1}}\Delta k_{L_{1}}}\\
l_{11} & = & \frac{e^{izk_{S}}\left(\left(\Delta k_{S_{2}}-\Delta k_{L_{2}}\right)e^{-iz\Delta k_{S_{2}}}-\Delta k_{S_{2}}e^{-iz\left(\Delta k_{S_{2}}-\Delta k_{L_{2}}\right)}+\Delta k_{L_{2}}\right)\varGamma_{2}\varLambda_{2}}{\left(\Delta k_{S_{2}}-\Delta k_{L_{2}}\right)\Delta k_{S_{2}}\Delta k_{L_{2}}}\\
l_{12} & = & \frac{e^{izk_{S}}\left(\left(\Delta k_{S_{1}}-\Delta k_{A_{1}}\right)e^{-iz\Delta k_{S_{1}}}-\Delta k_{S_{1}}e^{-iz\left(\Delta k_{S_{1}}-\Delta k_{A_{1}}\right)}+\Delta k_{A_{1}}\right)\varLambda_{1}\varOmega_{1}}{\left(\Delta k_{S_{1}}-\Delta k_{A_{1}}\right)\Delta k_{S_{1}}\Delta k_{A_{1}}}\\
l_{13} & = & \frac{e^{izk_{S}}\left(\left(\Delta k_{S_{1}}-\Delta k_{L_{1}}\right)e^{-iz\Delta k_{S_{2}}}-\Delta k_{S_{2}}e^{-iz\left(\Delta k_{S_{1}}-\Delta k_{L_{1}}\right)}+\Delta k_{A_{2}}\right)\varLambda_{2}\varOmega_{2}}{\left(\Delta k_{S_{1}}-\Delta k_{L_{1}}\right)\Delta k_{S_{2}}\Delta k_{A_{2}}}\\
l_{14} & = & -l_{15}=-l_{16}=-\frac{e^{izk_{S}}\left(e^{-iz\Delta k_{S}}+i\Delta k_{S}z-1\right)g^{2}}{\left(\Delta k_{S}\right)^{2}}\\
l_{17} & = & \frac{e^{izk_{S}}\left(e^{-iz\Delta k_{S_{1}}}+iz\Delta k_{S_{1}}-1\right)\varLambda_{1}^{2}}{\left(\Delta k_{S_{1}}\right)^{2}}\\
l_{18} & = & \frac{e^{izk_{S}}\left(e^{-iz\Delta k_{S_{2}}}+iz\Delta k_{S_{2}}-1\right)\varLambda_{2}^{2}}{\left(\Delta k_{S_{2}}\right)^{2}}
\end{array}
\end{equation}

\begin{equation}
\begin{array}{lcl}
m_{1} & = & e^{izk_{V}}\\
m_{2} & = & \frac{e^{izk_{V}}\left(1-e^{-iz\Delta k_{S}}\right)g}{\Delta k_{S}}\\
m_{3} & = & \frac{e^{izk_{V}}\left(1-e^{-iz\Delta k_{A}}\right)\chi}{\Delta k_{A}}\\
m_{4} & = & {{m_{5}=}{ m_{6}}{ =\frac{{ {g\chi e^{izk_{V}}\left(\left(\Delta k_{S}+\Delta k_{A}\right)\left(e^{-iz\Delta k_{A}}-e^{-iz\Delta k_{S}}\right)+\Delta k_{S}\left(e^{-iz\left(\Delta k_{S}+\Delta k_{A}\right)}-1\right)-\Delta k_{A}\left(e^{-iz\left(\Delta k_{S}+\Delta k_{A}\right)}-1\right)\right)}}}{{\left(\Delta k_{S}+\Delta k_{A}\right)\Delta k_{S}\Delta k_{A}}}}}\\
m_{7} & = & -\frac{g\Gamma_{1}e^{izk_{V}}\left(\left(\Delta k_{S}+\Delta k_{L_{1}}\right)e^{-iz\Delta k_{S}}-\Delta k_{S}e^{-iz\left(\Delta k_{S}+\Delta k_{L_{1}}\right)}-\Delta k_{L_{1}}\right)}{\left(\Delta k_{S}+\Delta k_{L_{1}}\right)\Delta k_{S}\Delta k_{L_{1}}}\\
m_{8} & =- & \frac{g\Gamma_{2}e^{izk_{V}}\left(\left(\Delta k_{S}+\Delta k_{L_{2}}\right)e^{-iz\Delta k_{S}}-\Delta k_{S}e^{-iz\left(\Delta k_{S}+\Delta k_{L_{2}}\right)}-\Delta k_{L2}\right)}{\left(\Delta k_{S}+\Delta k_{L_{2}}\right)\Delta k_{S}\Delta k_{L_{2}}}\\
m_{9} & = & -\frac{g\varLambda_{1}e^{izk_{V}}\left(\left(\Delta k_{S}-\Delta k_{S_{1}}\right)e^{-iz\Delta k_{S}}-\Delta k_{S}e^{-iz\left(\Delta k_{S}-\Delta k_{S_{1}}\right)}+\Delta k_{S_{1}}\right)}{\left(\Delta k_{S}-\Delta k_{S_{1}}\right)\Delta k_{S}\Delta k_{S_{1}}}\\
m_{10} & = & -\frac{g\varLambda_{2}e^{izk_{V}}\left(\left(\Delta k_{S}-\Delta k_{S_{2}}\right)e^{-iz\Delta k_{S}}-\Delta k_{S}e^{-iz\left(\Delta k_{S}-\Delta k_{S_{2}}\right)}+\Delta k_{S_{2}}\right)}{\left(\Delta k_{S}-\Delta k_{S_{2}}\right)\Delta k_{S}\Delta k_{S_{2}}}\\
m_{11} & = & \frac{\chi\Gamma_{1}e^{izk_{V}}\left(\left(\Delta k_{L_{1}}-\Delta k_{A}\right)e^{-iz\Delta k_{A}}+\Delta k_{A}e^{-iz\left(\Delta k_{A}-\Delta k_{L_{1}}\right)}-\Delta k_{L_{1}}\right)}{\left(\Delta k_{A}-\Delta k_{L_{1}}\right)\Delta k_{A}\Delta k_{L_{1}}}\\
m_{12} & = & \frac{\chi\Gamma_{2}e^{izk_{V}}\left(\left(\Delta k_{L_{2}}-\Delta k_{A}\right)e^{-iz\Delta k_{A}}+\Delta k_{A}e^{-iz\left(\Delta k_{A}-\Delta k_{L_{2}}\right)}-\Delta k_{L_{2}}\right)}{\Delta k_{A}\left(\Delta k_{A}-\Delta k_{L_{2}}\right)\Delta k_{L_{2}}}\\
m_{13} & = & -\frac{\chi\varOmega_{1}e^{izk_{V}}\left(\left(\Delta k_{A}+\Delta k_{A_{1}}\right)e^{-iz\Delta k_{A}}-\Delta k_{A}e^{-iz\left(\Delta k_{A}+\Delta k_{A_{1}}\right)}-\Delta k_{A_{1}}\right)}{\left(\Delta k_{A}+\Delta k_{A_{1}}\right)\Delta k_{A}\Delta k_{A_{1}}}\\
m_{14} & = & -\frac{\chi\varOmega_{2}e^{izk_{V}}\left(\left(\Delta k_{A}+\Delta k_{A_{2}}\right)e^{-iz\Delta k_{A}}-\Delta k_{A}e^{-iz\left(\Delta k_{A}+\Delta k_{A_{2}}\right)}-\Delta k_{A_{2}}\right)}{\left(\Delta k_{A}+\Delta k_{A_{2}}\right)\Delta k_{A}\Delta k_{A_{2}}}\\
m_{15} & =- & m_{16}=-m_{17}=-\frac{e^{izk_{V}}g^{2}\left(e^{-iz\Delta k_{S}}+i\Delta k_{S}z-1\right)}{\left(\Delta k_{S}\right)^{2}}\\
m_{18} & = & -m_{19}=-m_{20}=\frac{e^{izk_{V}}\chi^{2}\left(e^{-iz\Delta k_{A}}+i\Delta k_{A}z-1\right)}{\left(\Delta k_{A}\right)^{2}}
\end{array}
\end{equation}

\begin{equation}
\begin{array}{lcl}
n_{1} & = & e^{izk_{A}}\\
n_{2} & = & \frac{e^{izk_{A}}\left(-1+e^{iz\Delta k_{A}}\right)\chi}{\Delta k_{A}}\\
n_{3} & = & \frac{e^{izk_{A}}\left(1-e^{-iz\Delta k_{A_{1}}}\right)\varOmega_{1}}{\Delta k_{A_{1}}}\\
n_{4} & = & \frac{e^{izk_{A}}\left(1-e^{-iz\Delta k_{A_{2}}}\right)\varOmega_{2}}{\Delta k_{A_{2}}}\\
n_{5} & = & \frac{g\chi e^{izk_{A}}\left(\left(\Delta k_{S}-\Delta k_{A}\right)e^{iz\Delta k_{A}}+\Delta k_{A}e^{-iz\left(\Delta k_{S}-\Delta k_{A}\right)}-\Delta k_{S}\right)}{\left(\Delta k_{S}-\Delta k_{A}\right)\Delta k_{A}\Delta k_{S}}\\
n_{6} & = & n_{7}=-\frac{g\chi e^{izk_{A}}\left(\left(\Delta k_{S}+\Delta k_{A}\right)e^{iz\Delta k_{A}}-\Delta k_{A}e^{iz\left(\Delta k_{S}+\Delta k_{A}\right)}-\Delta k_{S}\right)}{\left(\Delta k_{S}+\Delta k_{A}\right)\Delta k_{A}\Delta k_{S}}\\
n_{8} & = & \frac{\chi\Gamma_{1}e^{izk_{A}}\left(\left(\Delta k_{A}-\Delta k_{L_{1}}\right)e^{iz\Delta k_{A}}-\Delta k_{A}e^{iz\left(\Delta k_{A}-\Delta k_{L_{1}}\right)}+\Delta k_{L_{1}}\right)}{\left(\Delta k_{A}-\Delta k_{L_{1}}\right)\Delta k_{A}\Delta k_{L_{1}}}\\
n_{9} & = & \frac{\chi\Gamma_{2}e^{izk_{A}}\left(\left(\Delta k_{A}-\Delta k_{L_{2}}\right)e^{iz\Delta k_{A}}-\Delta k_{A}e^{iz\left(\Delta k_{A}-\Delta k_{L_{2}}\right)}+\Delta k_{L_{2}}\right)}{\left(\Delta k_{A}-\Delta k_{L_{2}}\right)\Delta k_{A}\Delta k_{L_{2}}}\\
n_{10} & = & \frac{\varGamma_{1}\varOmega_{1}e^{izk_{A}}\left(\Delta k_{L_{1}}+\left(\Delta k_{A_{1}}-\Delta k_{L_{1}}\right)e^{-iz\Delta k_{A_{1}}}-\Delta k_{A_{1}}e^{-iz\left(\Delta k_{A_{1}}-\Delta k_{L_{1}}\right)}\right)}{\Delta k_{L_{1}}\left(\Delta k_{A_{1}}-\Delta k_{L_{1}}\right)\Delta k_{A_{1}}}\\
n_{11} & = & \frac{\varGamma_{2}\varOmega_{2}e^{izk_{A}}\left(\Delta k_{L_{2}}+\left(\Delta k_{A_{2}}-\Delta k_{L_{2}}\right)e^{-iz\Delta k_{A_{2}}}-\Delta k_{A_{2}}e^{-iz\left(\Delta k_{A_{2}}-\Delta k_{L_{2}}\right)}\right)}{\Delta k_{L_{2}}\left(\Delta k_{A_{2}}-\Delta k_{L_{2}}\right)\Delta k_{A_{2}}}\\
n_{12} & = & \frac{e^{izk_{A}}\left(e^{-iz\Delta k_{A_{1}}}\left(\Delta k_{A_{1}}-\Delta k_{S_{1}}\right)+\Delta k_{S_{1}}-e^{-iz\left(\Delta k_{A_{1}}-\Delta k_{S_{1}}\right)}\Delta k_{A_{1}}\right)\varLambda_{1}\varOmega_{1}}{\left(\Delta k_{A_{1}}-\Delta k_{S_{1}}\right)\Delta k_{S_{1}}\Delta k_{A_{1}}}\\
n_{13} & = & \frac{e^{izk_{A}}\left(e^{-iz\Delta k_{A_{2}}}\left(\Delta k_{A_{2}}-\Delta k_{S_{2}}\right)+\Delta k_{S_{2}}-e^{-iz\left(\Delta k_{A_{2}}-\Delta k_{S_{2}}\right)}\Delta k_{A_{2}}\right)\varLambda_{2}\varOmega_{2}}{\left(\Delta k_{A_{2}}-\Delta k_{S_{2}}\right)\Delta k_{S_{2}}\Delta k_{A_{2}}}\\
n_{14} & = & n_{15}=n_{16}=\frac{e^{izk_{A}}\chi^{2}\left(e^{iz\Delta k_{A}}-i\Delta k_{A}z-1\right)}{\Delta k_{A}^{2}}\\
n_{17} & = & \frac{e^{izk_{A}}\left(e^{-iz\Delta k_{A_{1}}}+i\Delta k_{A_{1}}z-1\right)\varOmega_{1}^{2}}{\left(\Delta k_{A_{1}}\right)^{2}}\\
n_{18} & = & \frac{e^{izk_{A}}\left(e^{-iz\Delta k_{A_{2}}}+i\Delta k_{A_{2}}z-1\right)\varOmega_{2}^{2}}{\left(\Delta k_{A_{2}}\right)^{2}}
\end{array}
\end{equation}
where $\Delta k_{S}=(k_{S}+k_{V}-k_{L_{1}}-k_{L_{2}})$, $\Delta k_{A}=(k_{L_{1}}+k_{L_{2}}+k_{V}-k_{A})$,
$\Delta k_{L_{1}}=(k_{L_{1}}-k_{p_{1}}),$ $\Delta k_{L_{2}}=\left(k_{L_{2}}-k_{p_{2}}\right)$
$\Delta k_{S_{1}}=(k_{S}-k_{p_{1}})$, $\Delta k_{S_{2}}=(k_{S}-k_{p_{2}})$,
$\Delta k_{A_{2}}=(k_{A}-k_{p_{2}})$, and $\Delta k_{A_{1}}=(k_{A}-k_{p_{1}})$. 
\end{document}